\documentclass[useAMS,usenatbib]{mn2e}
\usepackage{graphicx}
\usepackage{subfigure}

\newcommand{\Mjup}{\mbox{M$_{\rm Jup}$}}

\newcommand{\gtsimeq}{\raisebox{-0.6ex}{$\,\stackrel
         {\raisebox{-.2ex}{$\textstyle >$}}{\sim}\,$}}

\title[HU Aquarii Revisited]{Revisiting the proposed planetary system 
orbiting the eclipsing polar HU Aquarii}

\author[R.A. Wittenmyer, J. Horner, J.P. Marshall, O. Butters, \& C.G. Tinney]{Robert A.~Wittenmyer$^{1}$\thanks{E-mail: rob@phys.unsw.edu.au 
(RW)}, J.~Horner$^{1}$, J.P. Marshall$^{2}$, O.W. Butters$^{3}$ and C.G.~Tinney$^{1}$\\ 
$^{1}$Department of Astrophysics and Optics, School of Physics, 
University of New South Wales, Sydney 2052, Australia\\ 
$^{2}$Departmento F\'isica Te\'orica, Facultad de Ciencias, Universidad 
Aut\'onoma de Madrid, Cantoblanco, 28049, Madrid, Espa\~na\\
$^{3}$Department of Physics and Astronomy, University of Leicester, Leicester, LE1 7RH, UK}
\begin{document}

\date{Accepted  Received ; in original form }

\pagerange{\pageref{firstpage}--\pageref{lastpage}} \pubyear{2011}

\maketitle

\label{firstpage}

\begin{abstract}

It has recently been proposed, on the basis of eclipse-timing data, that 
the eclipsing polar cataclysmic variable HU~Aquarii is host to at least 
two giant planets. However, that result has been called into question 
based upon the dynamical stability of the proposed planets. In this 
work, we present a detailed re-analysis of all eclipse timing 
data available for the HU~Aquarii system, making use of standard 
techniques used to fit orbits to radial-velocity data. We find that the 
eclipse timings can be used to obtain a two-planet solution that does 
not require the presence of additional bodies within the system. We then 
perform a highly detailed dynamical analysis of the proposed planetary 
system.  We show that the improved orbital parameters we have derived 
correspond to planets that are dynamically unstable on unfeasibly short 
timescales (of order 10$^4$ years or less).  Given these results, we 
discuss briefly how the observed signal might in fact be the result of 
the intrinsic properties of the eclipsing polar, rather than being 
evidence of dynamically improbable planets. Taken in concert, our 
results highlight the need for caution in interpreting such timing 
variations as being planetary in nature.


\end{abstract}

\begin{keywords}

binaries: close, binaries: eclipsing. stars: individual: HU Aqr, 
planetary systems, white dwarfs

\end{keywords}

\section{Introduction}

Cataclysmic variables (CVs) are interacting binary stars composed of a 
white dwarf primary and a Roche lobe filling M dwarf secondary.  In the 
case of HU~Aqr, an AM~Her class CV, the material being accreted by the 
primary from the secondary is channelled along an accretion stream by 
the white dwarf's magnetic field.  A comprehensive overview of these 
systems can be found in \cite{hellier01}.  Such systems are known to 
exhibit quasi-periodic variations in their photometry.  A number of 
factors can contribute to those variations, with well accepted causes 
including the level of activity of the secondary (star spots and 
associated stellar cycles) and also that star's shape.

A number of recent studies have suggested that certain CVs host 
planetary-mass companions.  As the postulated planetary companions orbit 
the central stars, they cause those stars to move back and forth as they 
orbit around the system's centre of mass.  As a result, the distance 
between the Earth and the host stars varies as a function of time, 
meaning that the light from the stars must sometimes travel further to 
reach us than at other times.  This effect results in measurable 
variations in the timing of mutual eclipse events between the two stars 
that can be measured from the Earth.  Using this method, planetary mass 
companions have recently been announced around the CVs UZ~For 
\citep{potter11}, NN~Ser \citep{b10}, DP~Leo \citep{qian10}, and HU~Aqr 
\citep{qian11}.  In each of these studies, the eclipse timings are first 
fitted to a linear ephemeris.  The residuals from this ephemeris (the 
$O-C$, or $Observed - Calculated$ timings) are then plotted and found to 
show further, higher-order variations.  These $O-C$ timings can then be 
fit with one or more superposed Keplerian orbits in a manner essentially 
identical to that employed in Doppler radial-velocity planet detection.  
In \citet{horner11}, we used the methodology of \citet{marshall10} to 
simulate the long-term dynamical stability of the two giant planets 
proposed as orbiting HU~Aqr \citep{qian11}.  We showed that the nominal 
2-planet solution was extremely unstable on short timescales ($\sim10^5$ 
yr), unless the outer planet orbited in a retrograde and coplanar sense 
relative to the inner (i.e.~with the two planetary orbits inclined by 
180 degrees to one another).  Given that such a configuration seems 
highly unlikely, we suggested that either the system is currently 
undergoing a dynamical rearrangement (also highly improbable given the 
$\gtsimeq 10^9$ yr age of the post-main-sequence primary), or that the 
system is significantly different from that proposed by \citet{qian11}.

In this work, we apply the standard statistical methods used by the 
radial-velocity planet search community to the timing data for HU~Aqr 
given by \citet{qian11}.  Section~2 briefly describes the observational 
data used for our analysis.  In Section~3, we detail the analysis 
methods applied to these data, and the resulting planetary system 
configurations implied.  In Section~4 we discuss the dynamical 
implications of our results, before exploring possible alternatives to 
the planet hypothesis in Section~5.

\section{Observational Data}

Following the discovery paper by \citet{qian11}, we make use of the same 
82 $O-C$ eclipse egress times, consisting of 72 data points from the 
literature \citep{schwarz09}, and 10 new timings presented by 
\citet{qian11}.  All eclipse timings have been fitted with a linear 
ephemeris given by

\begin{equation}
2449102.920257 + 0.0868294936E .
\end{equation}

\begin{figure}
\includegraphics[scale=0.4]{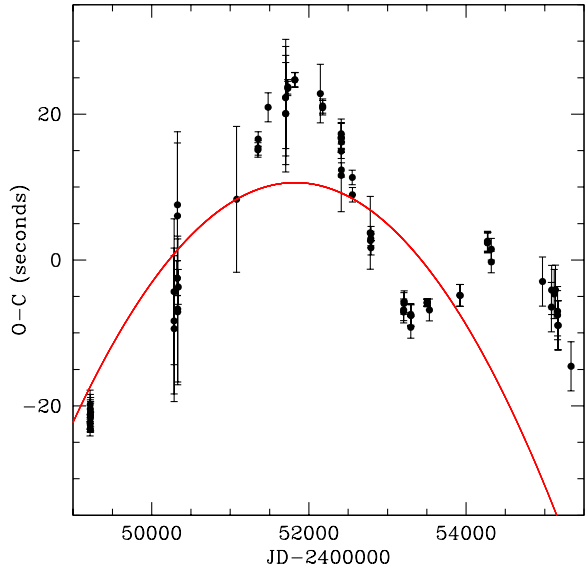}
\caption{$O-C$ eclipse timings for HU~Aqr, after fitting a linear 
ephemeris (given by Equation~1).  A quadratic fit to these data is 
overplotted as a solid line.  The data are those used in \citet{qian11}, 
which include 72 observations initially presented by \citet{schwarz09}.  
At least one sinusoidal variation is evident, suggesting the presence of 
at least one additional perturbing body. }
\label{inputdata}
\end{figure}

\noindent The residuals from the fit to this linear ephemeris are 
plotted in Figure~\ref{inputdata}.  The root-mean-square (RMS) scatter 
in the timing data is 13.4 seconds, and the reduced $\chi^2$ of the 
linear ephemeris is quite large at 108.6.  There are significant 
deviations from the linear ephemeris, suggestive of additional 
perturbing bodies which result in periodic eclipse timing variations.  
We therefore proceed with fitting Keplerian orbits to these signals.

\section{Orbit Fitting}

In this section, we detail the orbit-fitting process for two cases.  
First, in \S 3.1, we follow \cite{qian11}, and consider the data after 
the removal of a quadratic trend.  In \S 3.2, we attempt to fit the data 
without removing a quadratic trend.  It is well accepted that in fitting 
radial-velocity data (to which results on eclipse timing are clearly 
analogous), if a long-period object is suspected, the removal of a 
quadratic trend from the data is not ideal.  The physically meaningful 
function to fit and remove is a Keplerian (if the variation is thought 
to be due to planetary orbits).  As the Keplerian function is a complex 
one, and not necessarily well approximated by a quadratic, it is both 
preferable and more rigorous to attempt to fit a Keplerian orbit -- even 
if the parameters so derived are not well-constrained, one will at least 
not introduce spurious signals due to a poor match between a quadratic 
and a Keplerian.

\subsection{Removing a quadratic trend}

\begin{figure}
\includegraphics[scale=0.4]{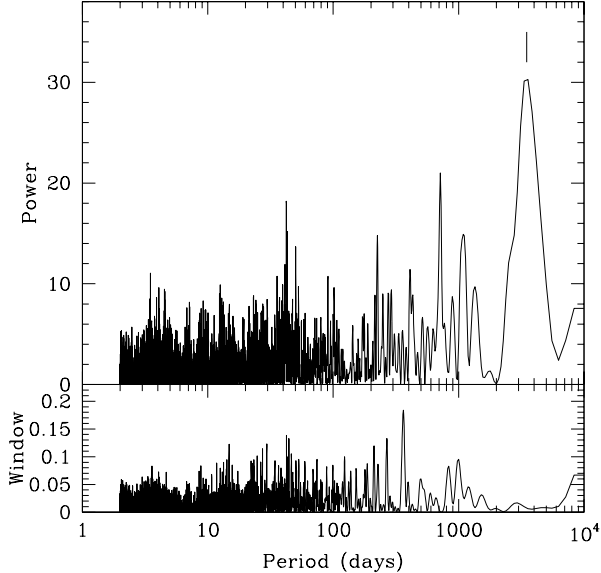}
\caption{Periodogram of the O-C timing data for HU~Aqr, using raw data 
(residuals from a linear ephemeris) with quadratic trend also removed. A 
strong signal is evident near 3500 days. }
\label{pgrams1}
\end{figure}

\begin{figure*}
\subfigure[]{
 \includegraphics[scale =0.35]{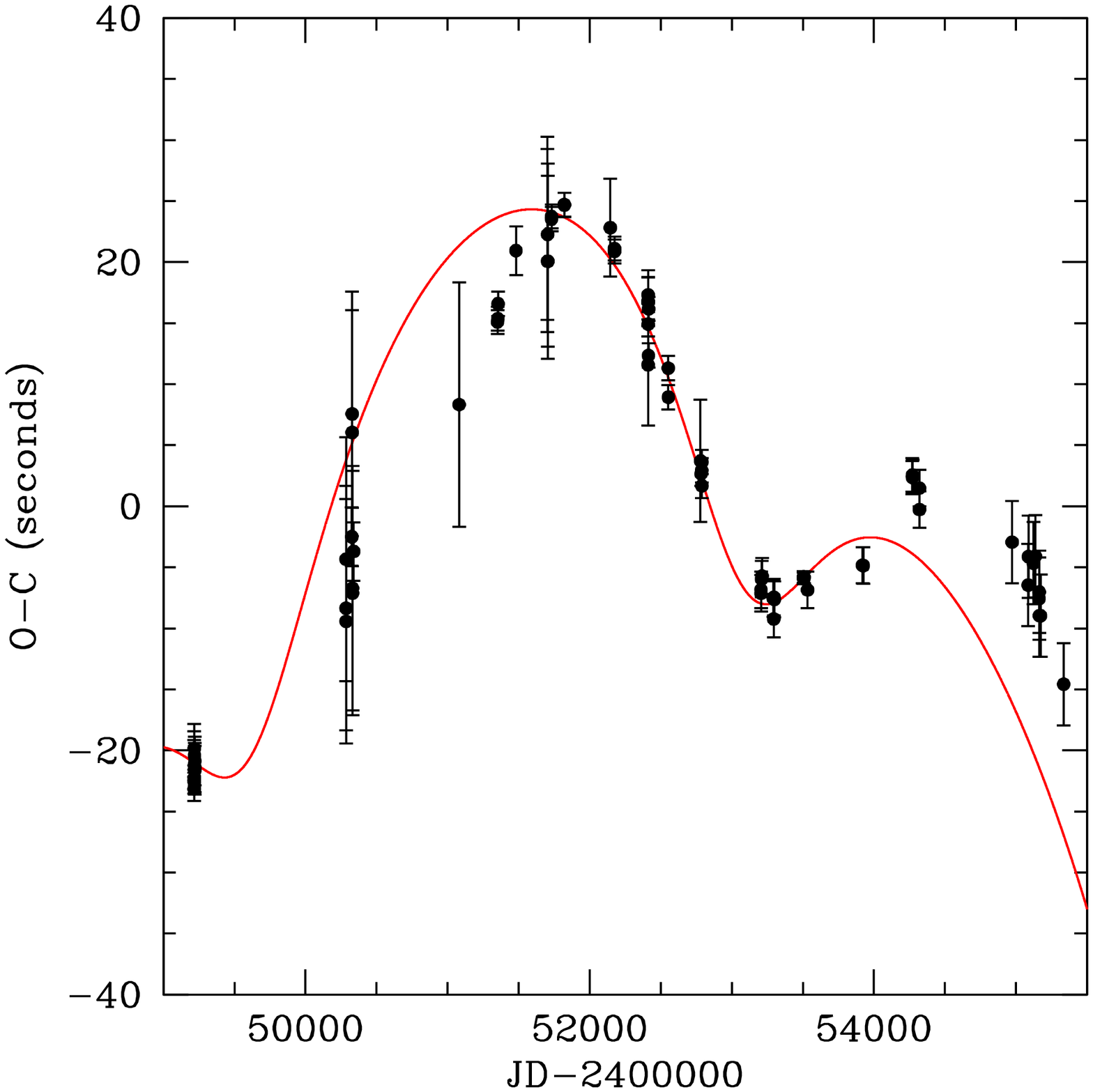}}
\subfigure[]{
 \includegraphics[scale =0.35]{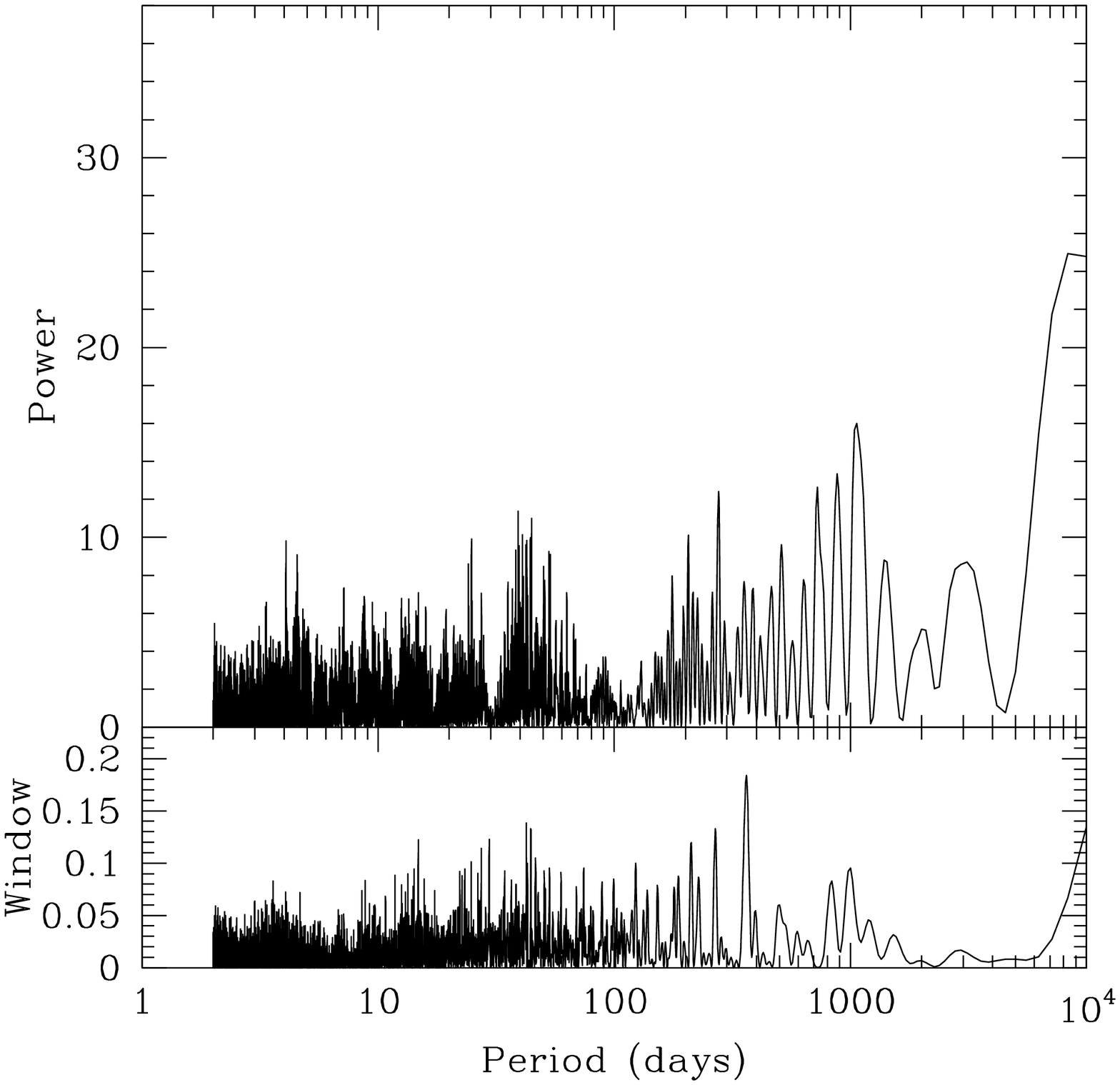}}
\caption{Results of single-planet fit on data with quadratic trend 
removed (Model~A1).  Left panel: Data and model fit for a single planet.  
The quadratic trend we have removed is superposed on the 1-planet 
Keplerian model (solid line).  Right panel: Periodogram of residuals 
after fitting for and removing the dominant signal, a Keplerian orbit 
with period 3538 days.  A significant peak is seen at a very long 
period. }
\label{A1resids}
\end{figure*}

In order to test the results of \cite{qian11}, we match their approach 
by fitting Keplerian orbits after removing a long-term quadratic trend.  
First, we fit a quadratic trend to the original timing data; the fitted 
parameters are

\begin{equation}
(O-C) = (-4.1\times 10^{-6})(JD)^2 + 0.43(JD) - 1.10\times 10^{4} ,
\end{equation}

\noindent where $JD$ is the observation date in the form (Julian 
Date-2400000).  Then, we fit a single planet (Model~A1).  A standard 
Lomb-Scargle periodogram \citep{lomb76, scargle82} shows a clear signal 
near 3500 days (Figure~\ref{pgrams1}).  We fit a Keplerian orbit model 
using \textit{GaussFit} \citep{jeffreys87}.  The single-planet solution 
is given in Table~\ref{planetparams} as ``Model~A1.'' The residuals to 
the 1-planet fit are shown in Figure~\ref{A1resids}, as is the 
periodogram of those residuals.  After removing the dominant 
periodicity, there is a significant peak at $\gtsimeq$8000 days.  Using 
the bootstrap randomisation method described by \citet{kurster97}, this 
peak is found to have a false-alarm probability (FAP) of $<$0.01\% 
(10000 bootstraps).  Owing to the high significance of this residual 
peak, and the large RMS of the 1-planet residuals (6.57 sec: 
Table~\ref{planetparams}), we proceed by fitting a second planet.

\begin{table}
  \centering
  \caption{Orbital Solutions for HU Aquarii}
  \begin{tabular}{llll}
  \hline
Model & Parameter & Inner Planet & Outer Planet \\
 \hline
A1 & Orbital Period (days)   & 3538$\pm$54 & \\
  & Amplitude (sec)          & 12.5$\pm$0.8 & \\
  & Eccentricity             & 0.26$\pm$0.05 & \\
  & $\omega$ (degrees)       & 171$\pm$12 & \\
  & $T_0$ (JD-2400000)       & 53092$\pm$517 & \\
  & Orbital Radius (AU)      & 4.66$\pm$0.11 & \\
  & $M\sin i$ (\Mjup)        & 6.08$\pm$0.13 & \\
 \hline
  & $\chi^2_{\nu}$            & 7.40 & \\
  & RMS (sec)                & 6.57 & \\
\hline
\hline
A2 & Orbital Period (days)   & 4647$\pm$36 & 7215$\pm$603 \\
  & Amplitude (sec)          & 20.2$\pm$0.5 & 23.4$\pm$1.9 \\
  & Eccentricity             & 0.19$\pm$0.02 & 0.53 (fixed) \\
  & $\omega$ (degrees)       & 166$\pm$5 & 190 (fixed) \\
  & $T_0$ (JD-2400000)       & 53074$\pm$50 & 58060$\pm$599 \\
  & Orbital Radius (AU)      & 5.59$\pm$0.11 & 7.50$\pm$0.50 \\
  & $M\sin i$ (\Mjup)        & 8.19$\pm$0.24 & 7.07$\pm$0.41 \\
 \hline
  & $\chi^2_{\nu}$            & 0.69 & \\
  & RMS (sec)                & 2.50 & \\
\hline
\hline
B1 & Orbital Period (days)   & $4728^{+300}_{-250}$ & \\
  & Amplitude (sec)          & 15.1$\pm$1.6 & \\
  & Eccentricity             & $0.09^{+0.05}_{0.09}$ & \\
  & $\omega$ (degrees)       & $181^{+55}_{-10}$ & \\
  & $T_0$ (JD-2400000)       & 53991$\pm$60 & \\
  & Orbital Radius (AU)      & 5.66$\pm$0.31 & \\
  & $M\sin i$ (\Mjup)        & 6.05$\pm$0.27 & \\
 \hline
  & $\chi^2_{\nu}$            & 21.2 & \\
  & RMS (sec)                & 8.00 & \\
\hline
\hline
B2 & Orbital Period (days)   & 4688$\pm$177 & 8377$\pm$610 \\
  & Amplitude (sec)          & 14.0$\pm$2.1 & 27.9$\pm$3.7 \\
  & Eccentricity             & 0.20$\pm$0.04 & $0.38^{+0.16}_{-0.11}$ (fixed) \\
  & $\omega$ (degrees)       & 254$\pm$12 & 332 (fixed) \\
  & $T_0$ (JD-2400000)       & 53640$\pm$116 & 60126$\pm$612 \\
  & Orbital Radius (AU)      & 5.62$\pm$0.22 & 8.28$\pm$0.50 \\
  & $M\sin i$ (\Mjup)        & 5.65$\pm$0.20 & 7.64$\pm$0.12 \\
 \hline
  & $\chi^2_{\nu}$            & 0.80 & \\
  & RMS (sec)                & 2.49 & \\
 \hline
 \end{tabular}
\label{planetparams}
\end{table}

\begin{figure*}
\subfigure[]{
 \includegraphics[scale=0.35]{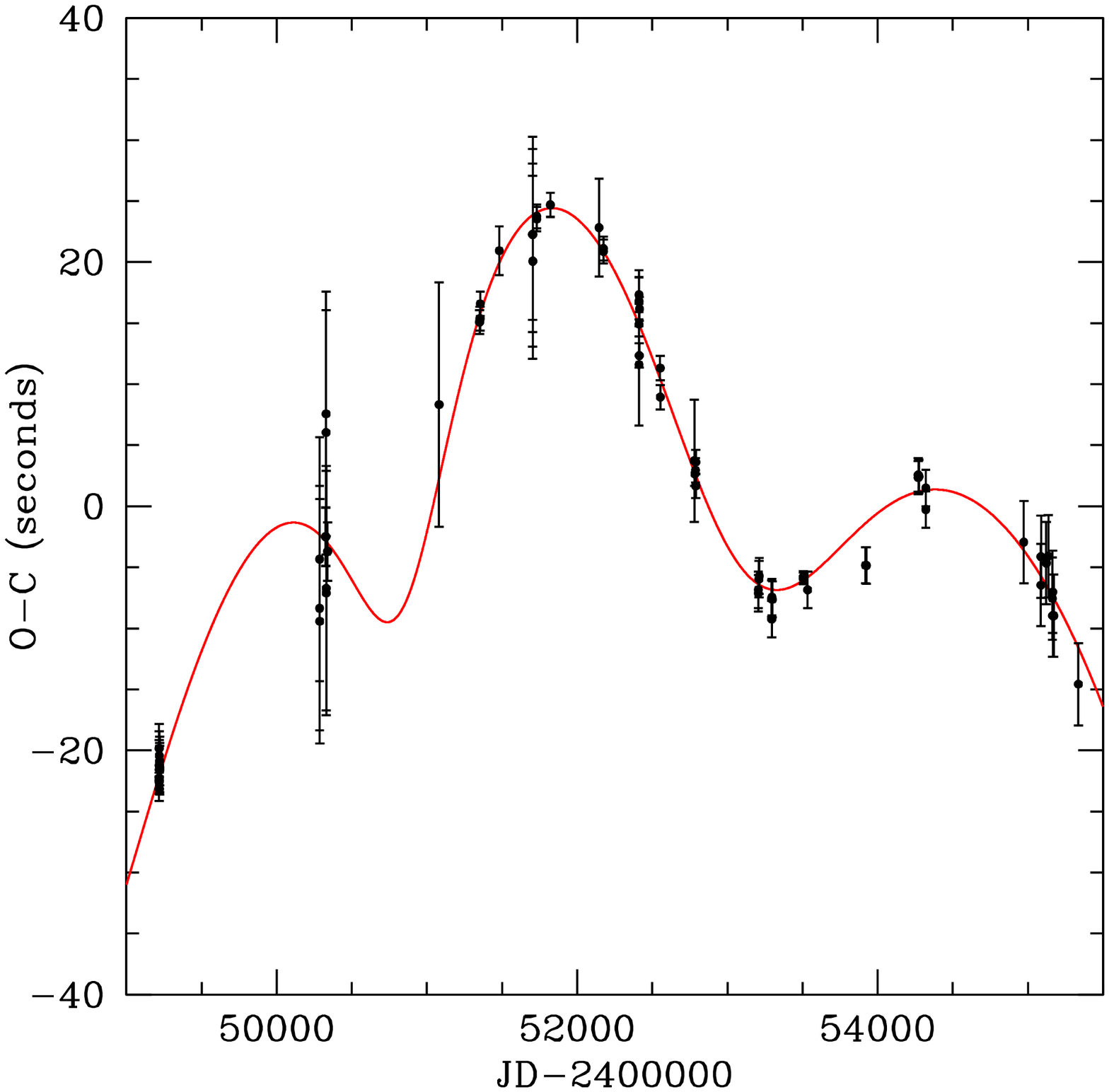}}
\subfigure[]{
 \includegraphics[scale=0.35]{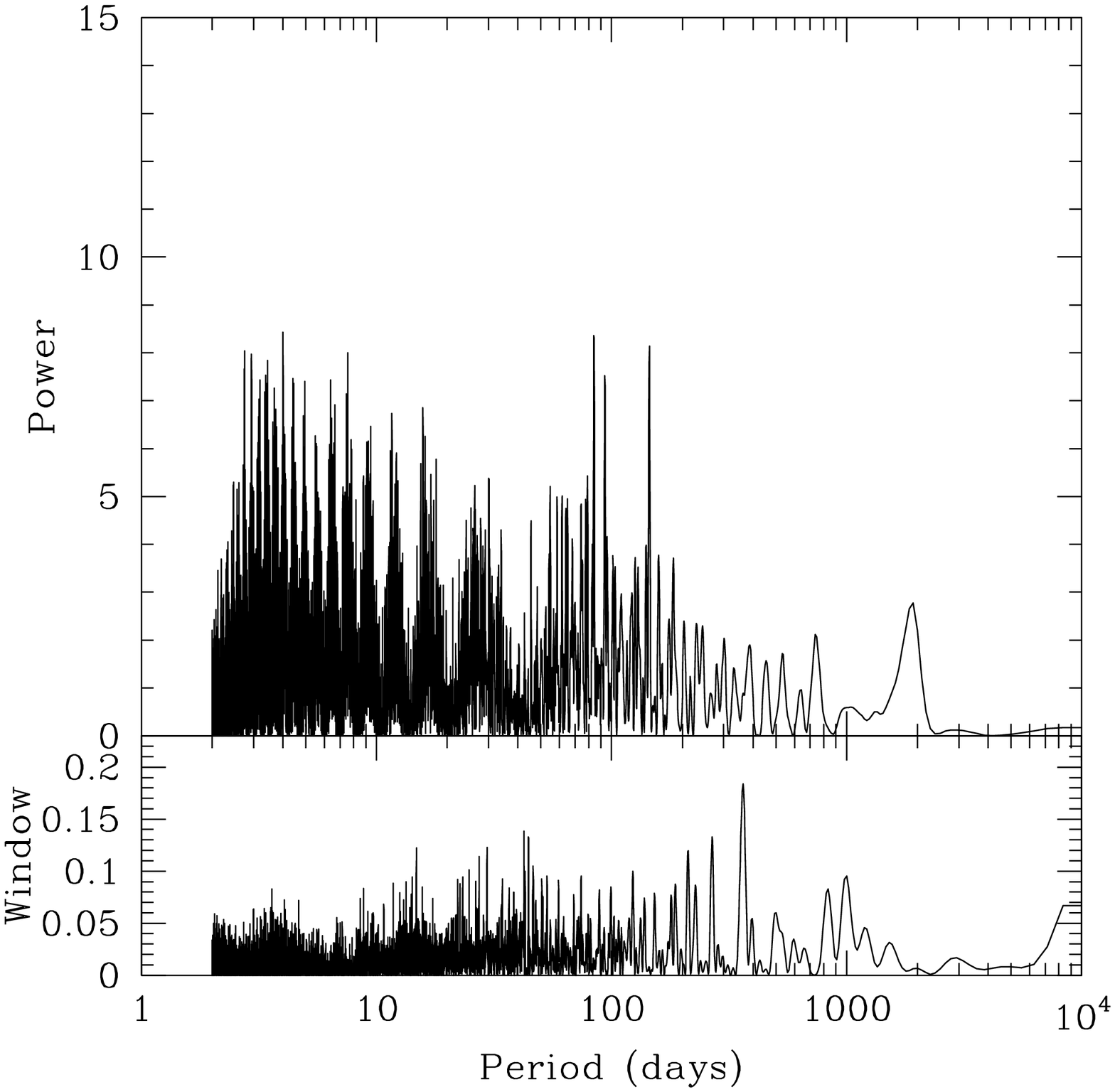}}
\caption{Left panel: Two-planet fit (Model~A2) on data with quadratic 
trend removed.  The quadratic trend we have removed is superposed on the 
2-planet Keplerian model (solid line).  Right panel: Periodogram of 
residuals to this fit.  No further significant periods are evident. }
\label{A2fits}
\end{figure*}

\begin{figure*}
\subfigure[]{
 \includegraphics[scale=0.35]{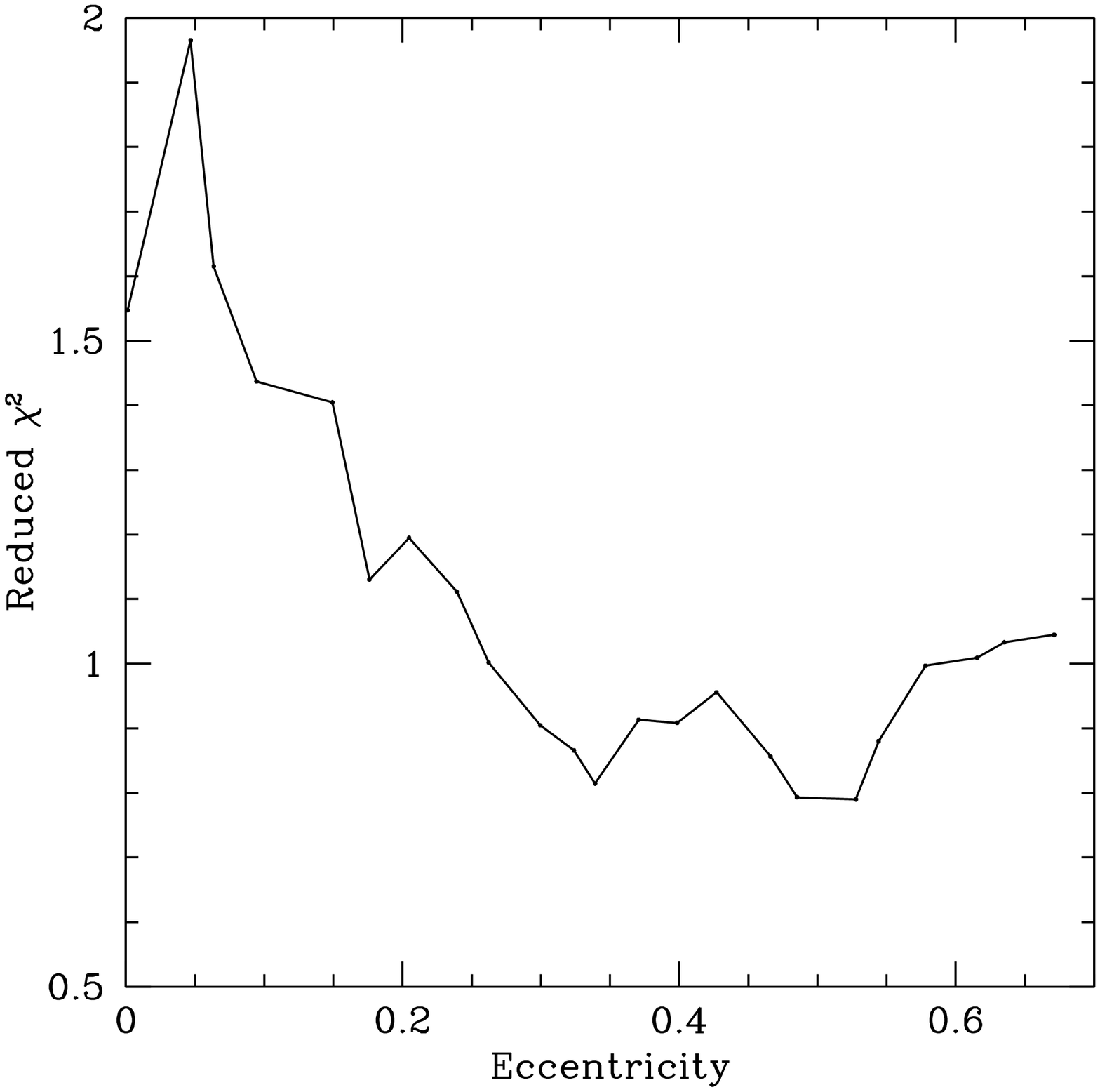}}
\subfigure[]{
 \includegraphics[scale=0.35]{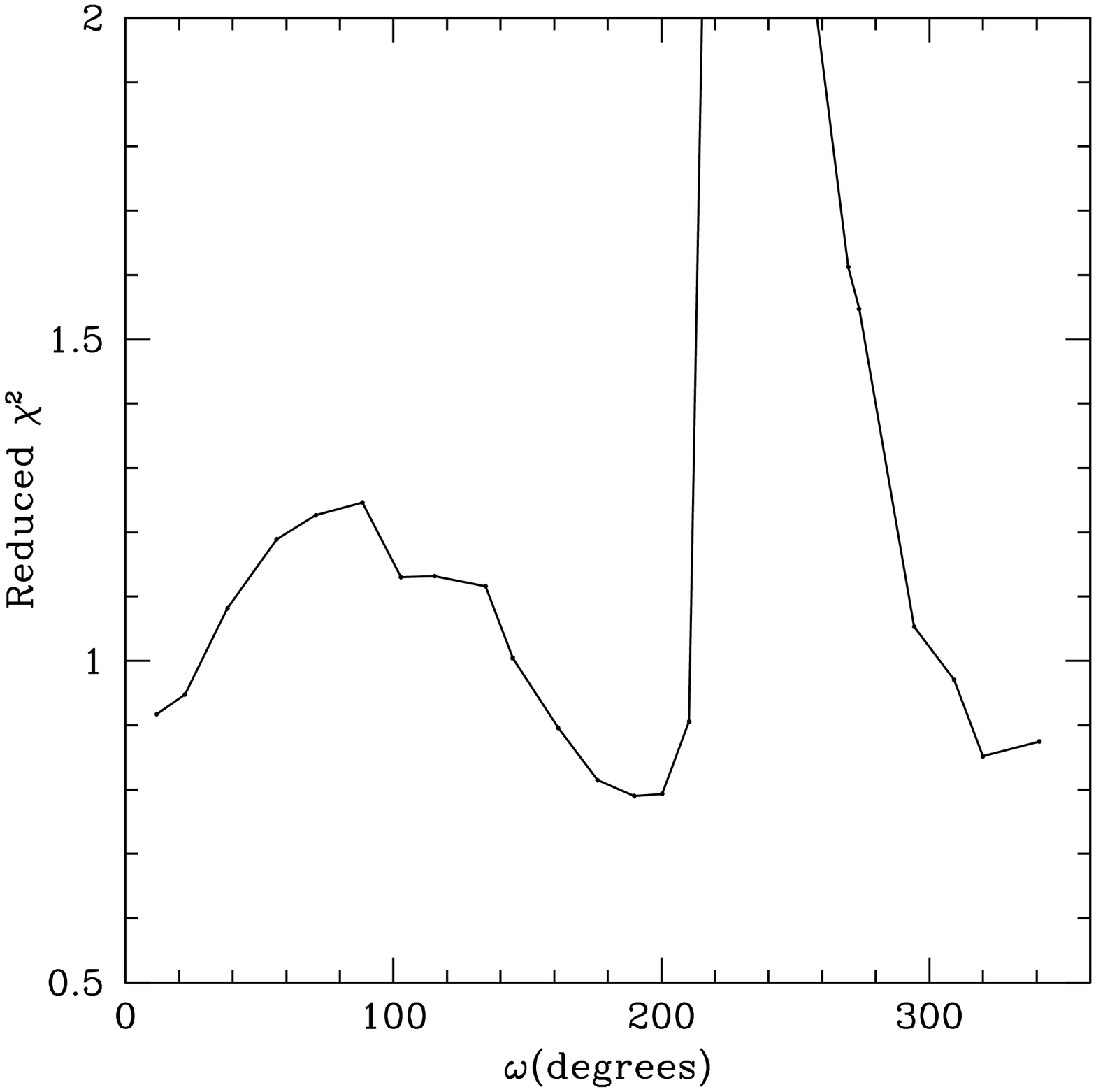}}
\caption{Results of genetic algorithm fitting for 2 planets (Model~A2).  
Left panel: Dependence of reduced $\chi^2$ on the outer planet's 
eccentricity.  Right panel: Same, but for the outer planet's argument of 
periastron ($\omega$). These parameters are essentially unconstrained, 
as nearly the entire allowed range is within 1.0 of the $\chi^2$ 
minimum. }
\label{wildandfree}
\end{figure*}

Given the substantial uncertainty in the system parameters, we used a 
genetic algorithm to explore a wide parameter space (e.g.~Cochran et 
al.~2007, Tinney et al.~2011).  The initial range of orbital periods 
supported by these data were first estimated by the periodogram analysis 
described above.  The parameters of the best 2-planet solution obtained 
by the genetic algorithm were then used as initial inputs for the 
\textit{GaussFit} least-squares fitting procedure used above.  The 
two-planet fit and a periodogram of its residuals are shown in 
Figure~\ref{A2fits}.  The parameters of the 2-planet fit are given in 
Table~\ref{planetparams} as ``Model~A2.''  Since the total duration of 
the data set is 6118 days, and the best-fit period for an outer body is 
7215 days, there remains significant uncertainty in the 2-planet fit.  
In particular, the Keplerian orbit-fitting process failed to converge 
when the outer planet's eccentricity and periastron argument ($\omega$) 
were allowed to be free parameters.  Hence, the values for these 
parameters shown in Table~\ref{planetparams} were held fixed at the best 
result from the genetic algorithm.  As shown in 
Figure~\ref{wildandfree}, these two parameters are almost completely 
unconstrained.  This is not surprising given that the period of the 
outer planet is larger than the entire length of the data set.

\begin{figure}
\includegraphics[scale=0.4]{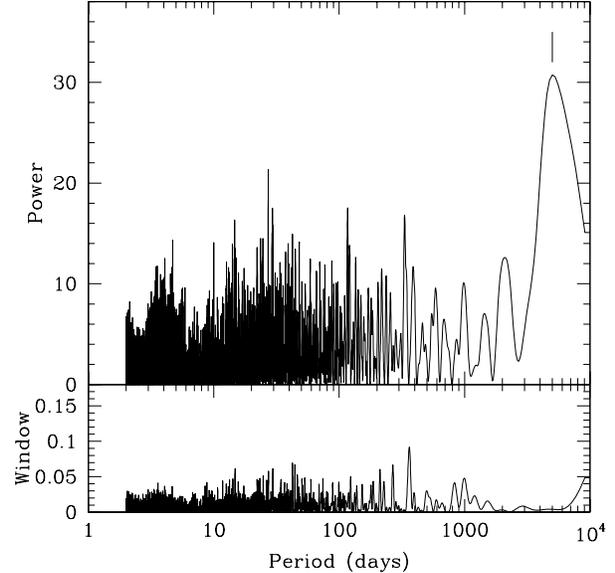}
\caption{Periodogram of the O-C timing data for HU~Aqr, using the raw 
data (residuals from a linear ephemeris) with \textit{no} additional 
quadratic trend removed. A strong signal is evident near 5000 days. }
\label{pgramsB1}
\end{figure}

\subsection{No quadratic trend}

\begin{figure*}
\subfigure[]{
 \includegraphics[scale=0.35]{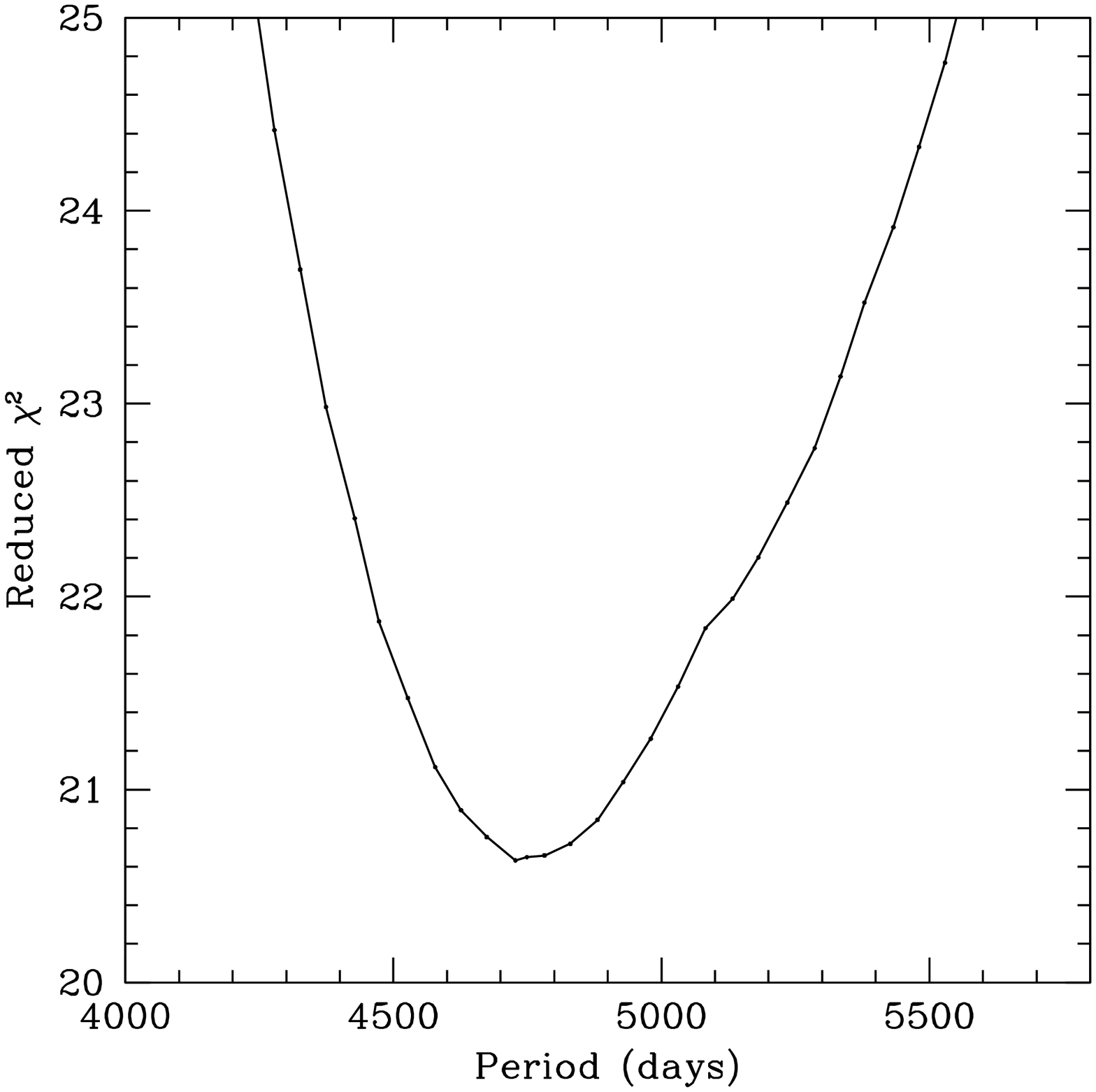}}
\subfigure[]{
 \includegraphics[scale=0.35]{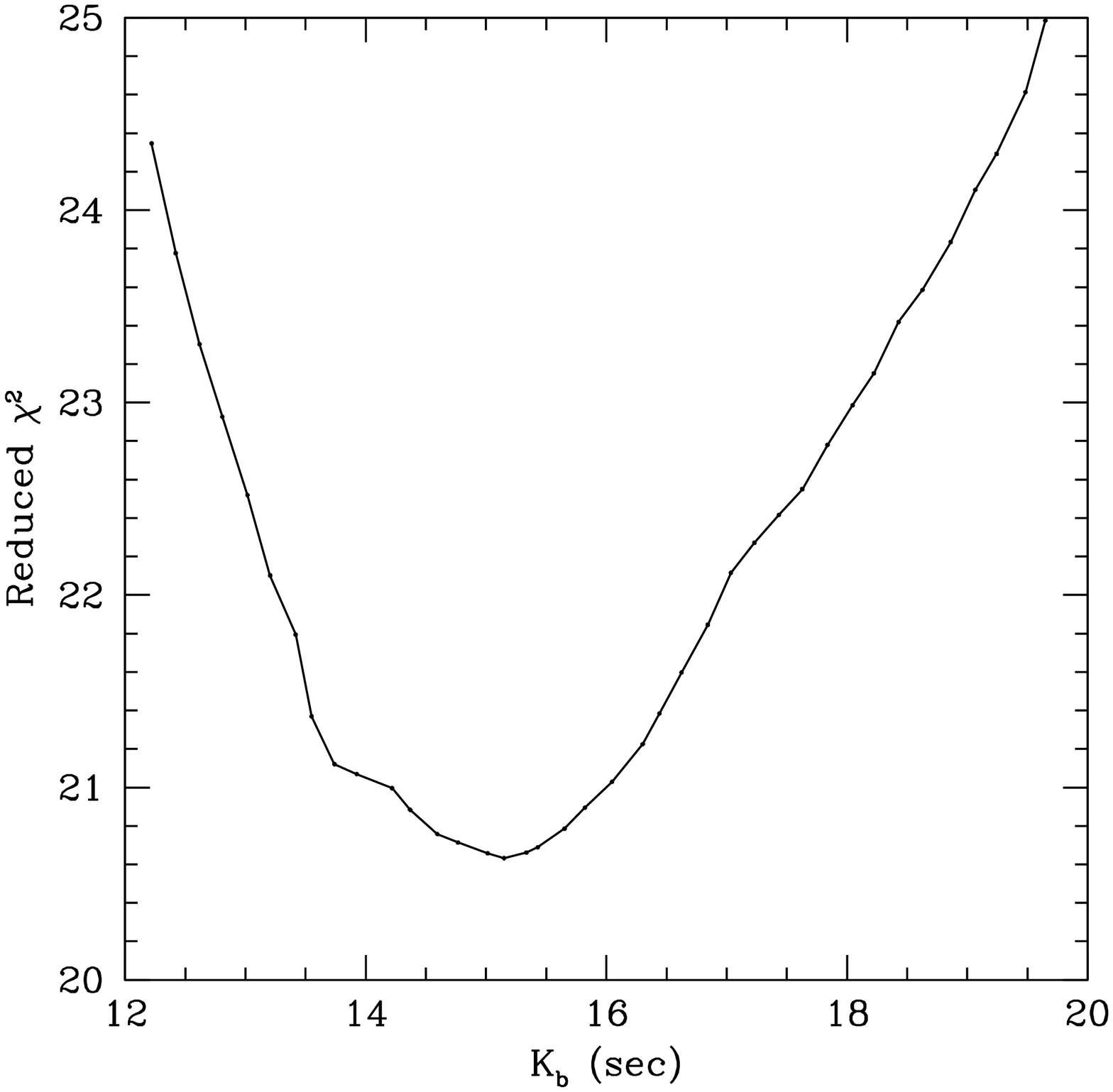}}
\subfigure[]{
 \includegraphics[scale=0.35]{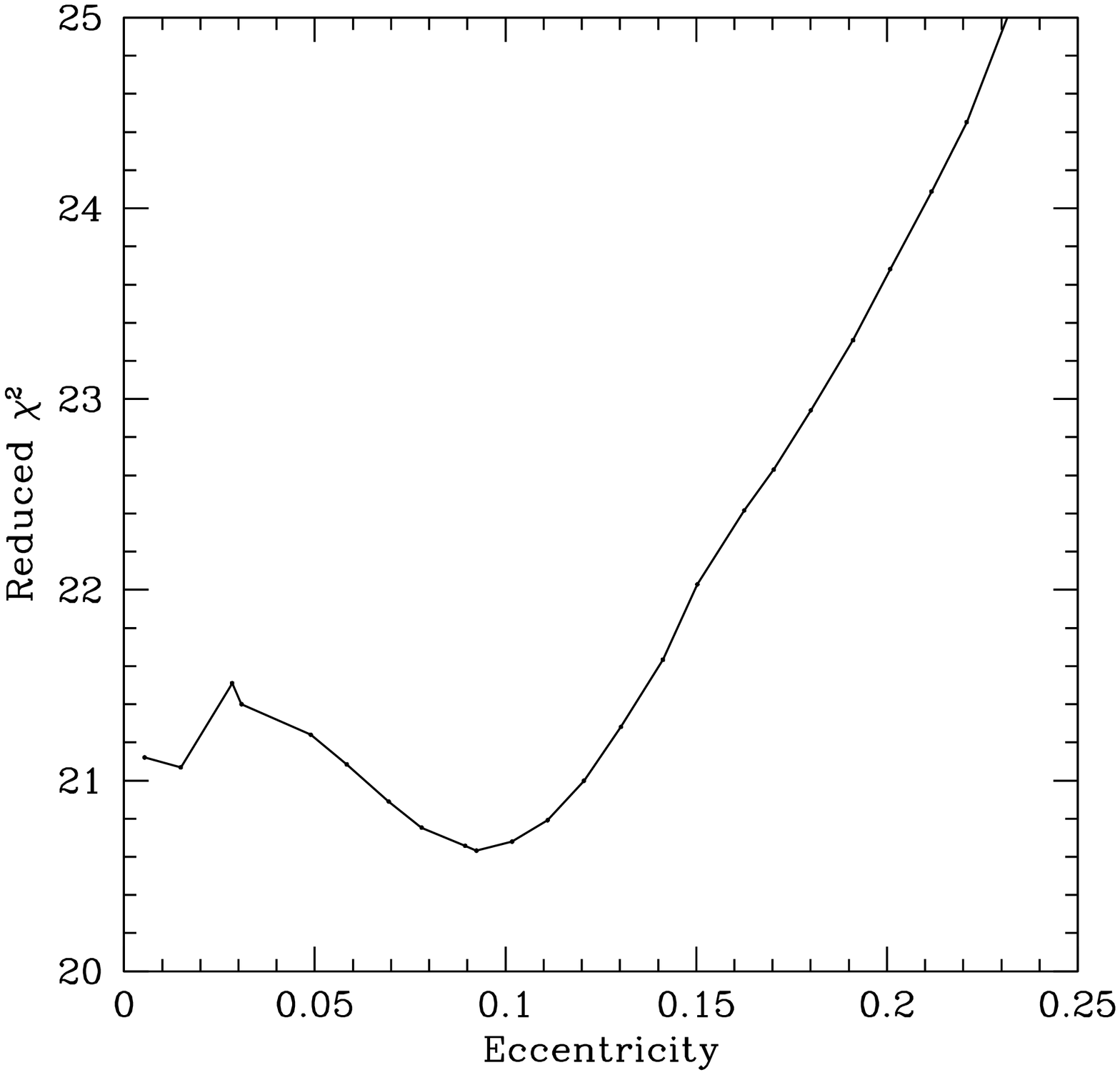}}
\subfigure[]{
 \includegraphics[scale=0.35]{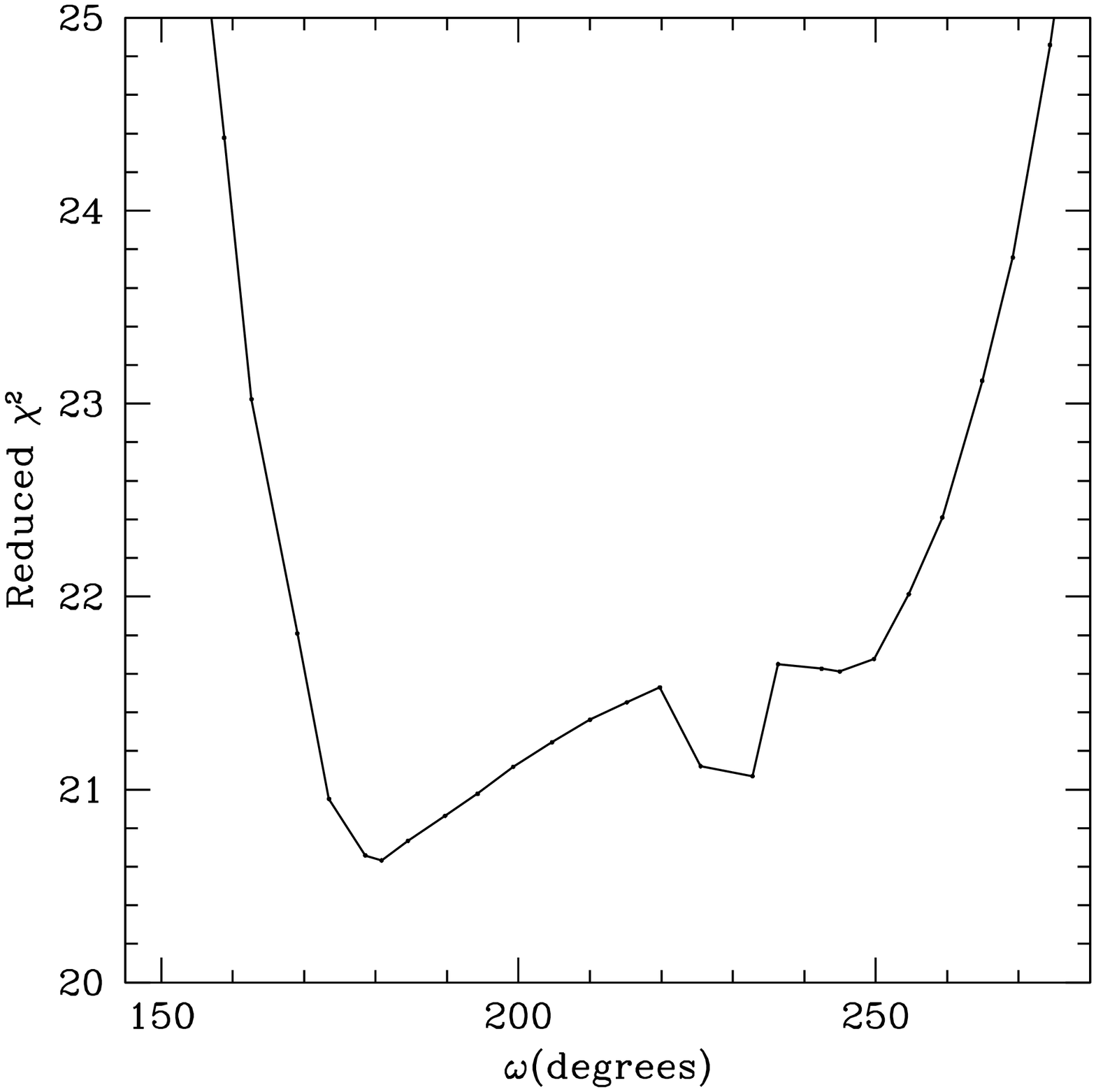}}
\caption{Results of single-planet genetic algorithm fit for HU~Aqr 
(Model~B1).  Each panel shows the dependence of reduced $\chi^2$ on a 
particular planetary parameter.  The uncertainty of each parameter is 
estimated by the range over which the reduced $\chi^2$ increases by 1 
from the minimum.}
\label{modelB1}
\end{figure*}

\begin{figure*}
\subfigure[]{
 \includegraphics[scale=0.35]{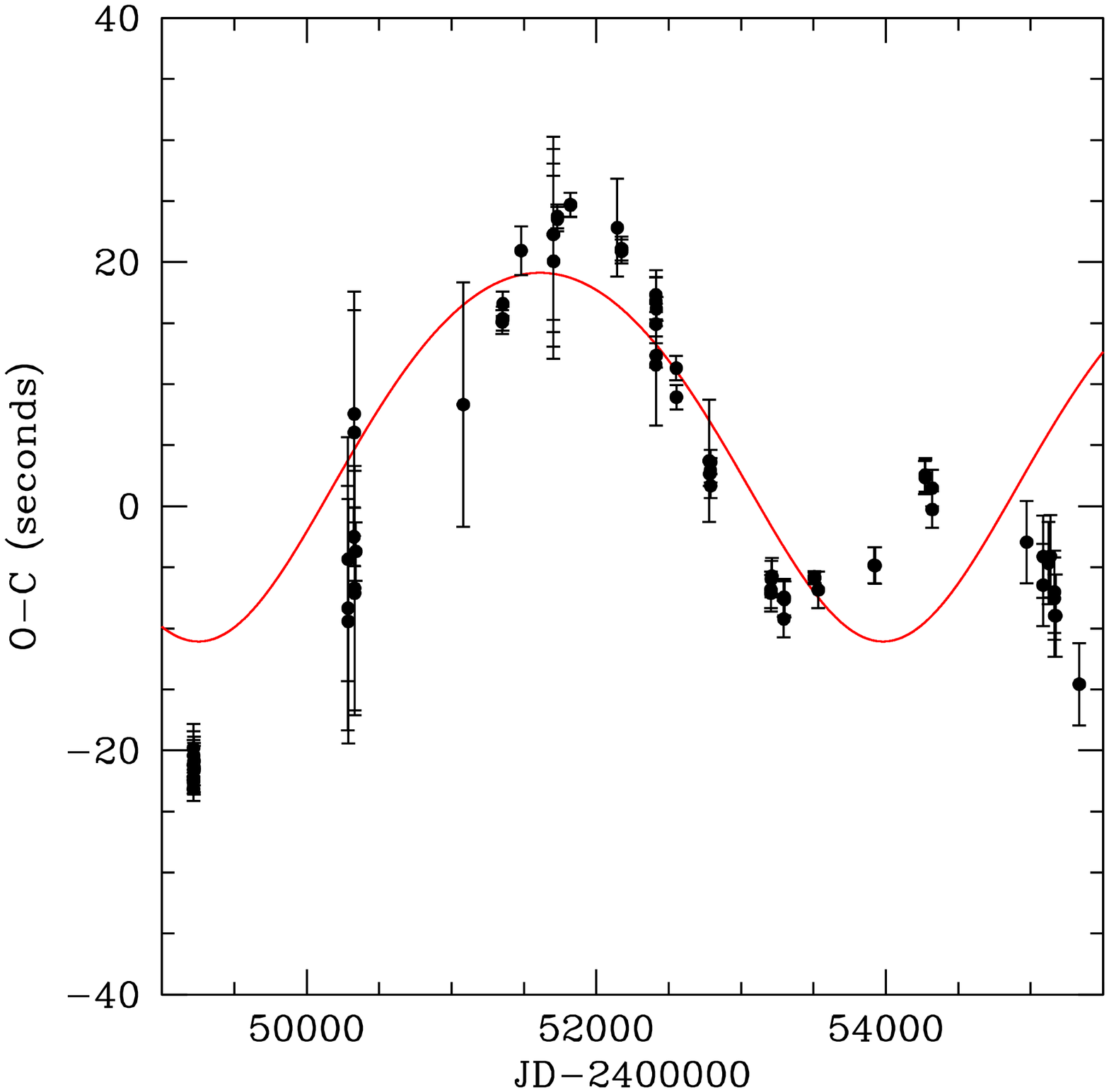}}
\subfigure[]{
 \includegraphics[scale=0.35]{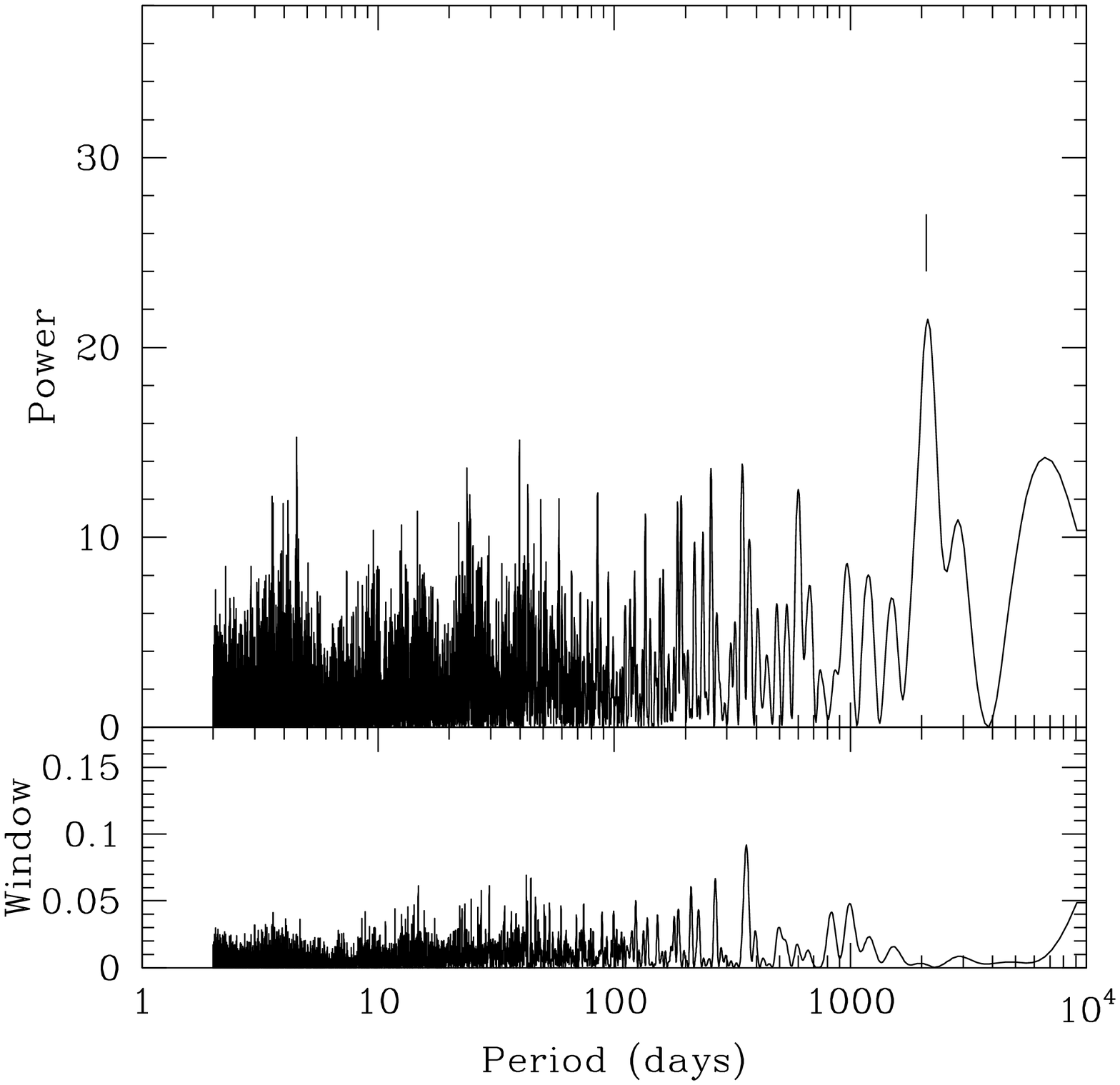}}
\caption{Results of single-planet fit for HU~Aqr (Model~B1).  Left 
panel: Data and model fit for a single planet.  Clearly a single planet 
is inadequate to fit these data.  Right panel: Periodogram of residuals 
after fitting for and removing the dominant signal, a Keplerian orbit 
with period 4728 days.  A significant peak is seen near 2100 days. }
\label{B1resids}
\end{figure*}

In this subsection, we explore the possibility that the removal of a 
quadratic trend has confounded the orbit-fitting process by absorbing 
signal due to a long-period outer planet.  Here we repeat the fitting 
procedures as above, but using the HU~Aqr data which have \textit{not} 
had a quadratic trend subtracted.  First, we considered a single planet 
by performing a periodogram search (Figure~\ref{pgramsB1}), which shows 
the dominant periodicity to be at 5000 days.  However, the standard 
approach of fitting a Keplerian orbit with \textit{GaussFit} failed.  As 
the reduced $\chi^2$ of the best-fit genetic algorithm result was an 
inordinately high 21.2, we attribute the failure of the least-squares 
method to the presence of additional signals.  In 
Table~\ref{planetparams}, we give the best-fit results for a single 
planet (``Model~B1'') from 100,000 iterations of the genetic algorithm.  
One-sigma uncertainties are estimated by noting the change in each 
parameter required to increase the reduced $\chi^2$ by 1.  The 
dependence of $\chi^2$ on each parameter is shown in 
Figure~\ref{modelB1}.  As with the one-planet solution in the previous 
trial (data which included a quadratic trend), the RMS scatter about a 
1-planet model is quite large at 8.0 seconds, compared to the mean 
measurement uncertainty of 2.4~seconds.  The fit and a periodogram of 
its residuals are shown in Figure~\ref{B1resids}; it is clear that one 
planet is not sufficient here.  The highest periodogram peak is at a 
period of 2128 days, again with a FAP$<$0.01\%.  We thus proceed to fit 
a second Keplerian orbit.


Again we employ the genetic algorithm to explore the wide and uncertain 
parameter space for a 2-planet model (``Model~B2'').  First we examine 
the short-period option, as prompted by the periodogram results in 
Figure~\ref{B1resids}.  We ran the genetic algorithm for 100,000 
iterations, then attempted a standard least-squares fit on the best 
result.  As with Model~B1, no such fit would converge with the 
eccentricities and periastron arguments of the two planets free.  The 
best-fit model with a short period for the second planet (actually 
making it the \textit{inner} of the two planets) resulted in a reduced 
$\chi^2$ of 4.06 and an RMS of 5.97 seconds.  The planetary parameters 
resulting from this fit are quite similar to those proposed by 
\citet{qian11}, with $P_{inner}=1947\pm$10 days and 
$P_{outer}=4429\pm$113 days.  However, this is substantially worse than 
the two-planet fit from Model~A2, and also worse than the long-period 
fit obtained below.

\begin{figure*}
\subfigure[]{
 \includegraphics[scale=0.35]{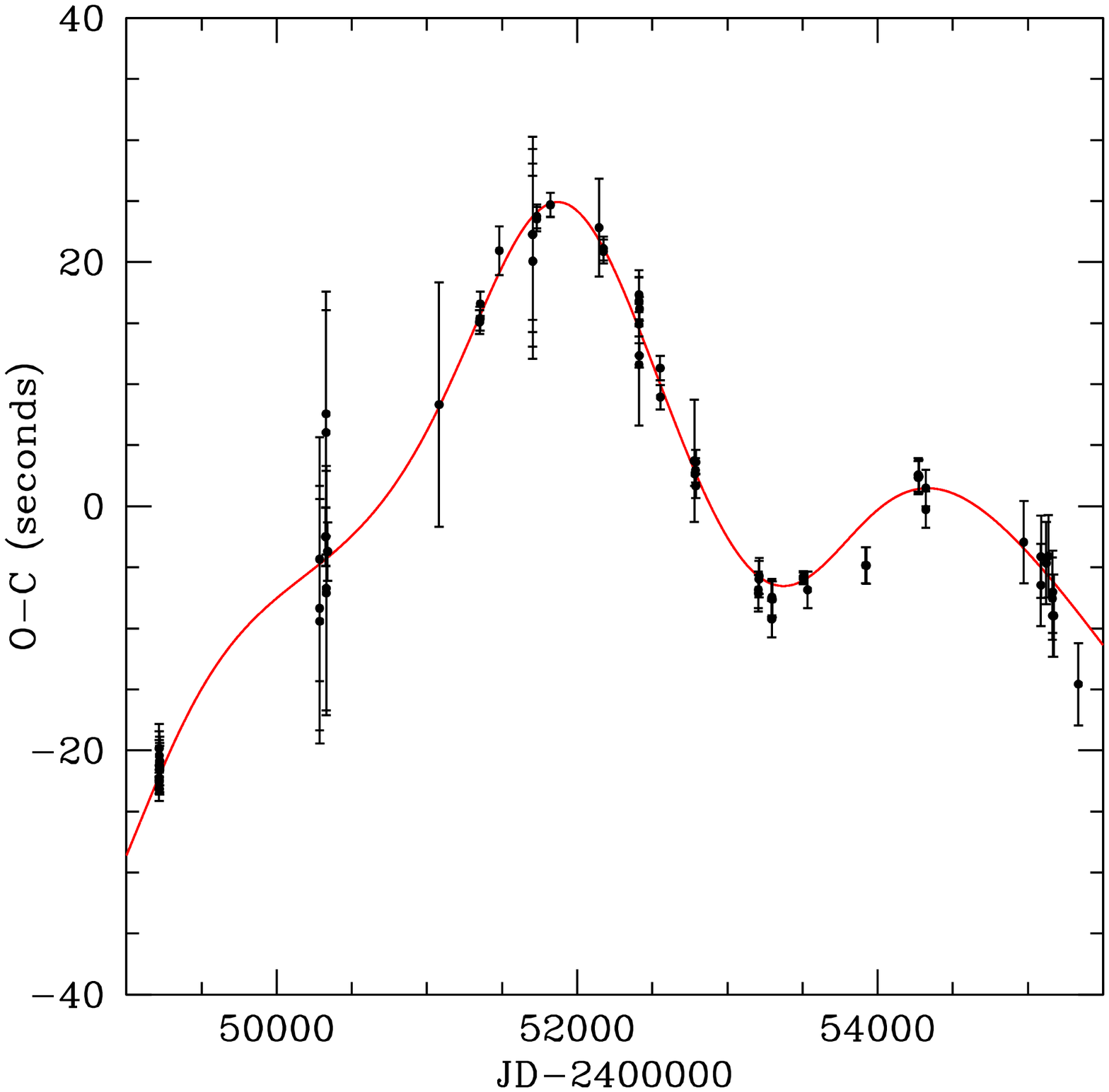}}
\subfigure[]{
 \includegraphics[scale=0.35]{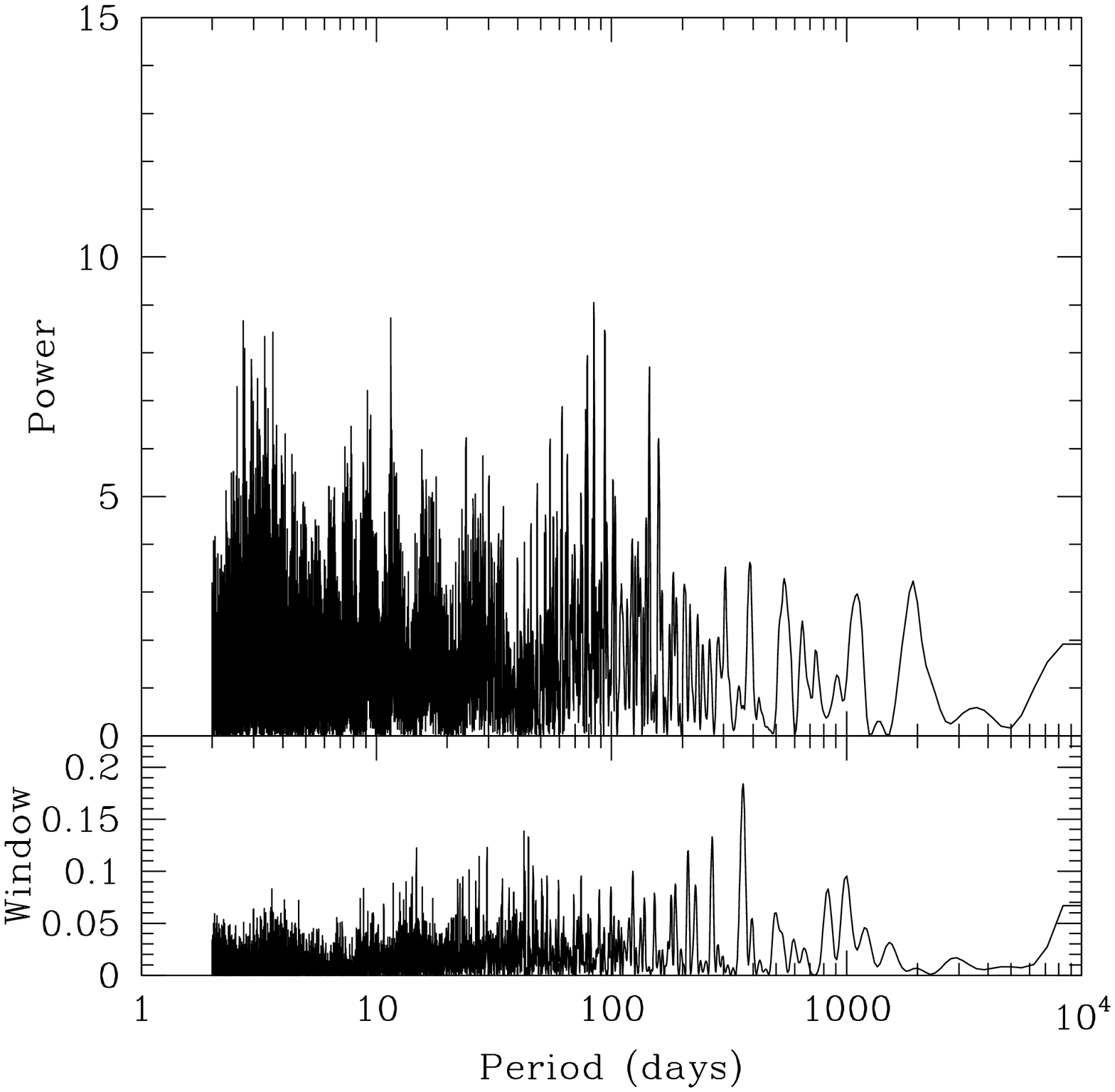}}
\caption{Left panel: Two-planet fit (Model~B2) to the observed data with 
no quadratic trend removed.  Right panel: Periodogram of residuals to 
this fit.  No further significant periods are evident. }
\label{B2fits}
\end{figure*}

\begin{figure*}
\subfigure[]{
 \includegraphics[scale =0.35]{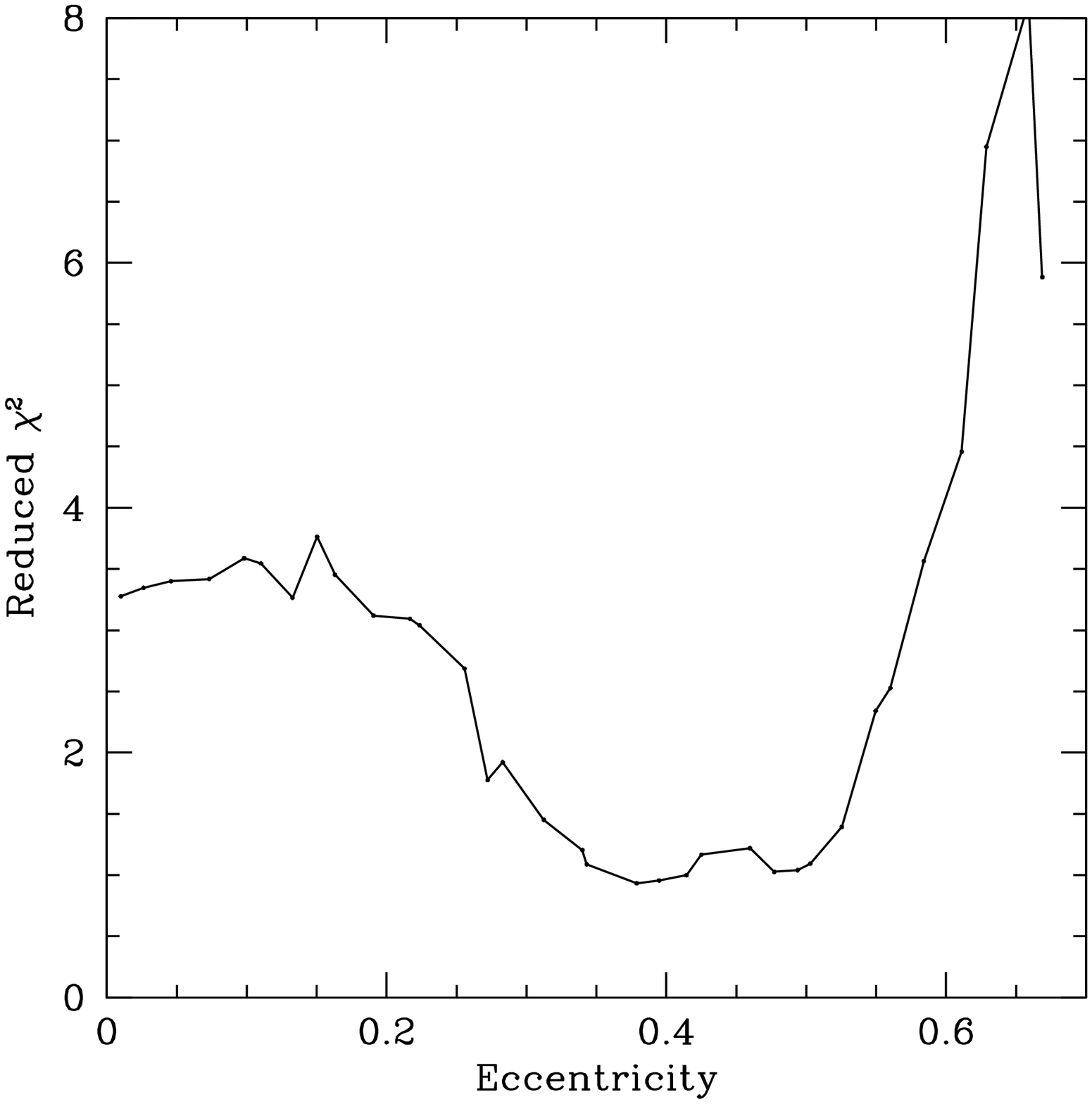}}
\subfigure[]{
 \includegraphics[scale =0.35]{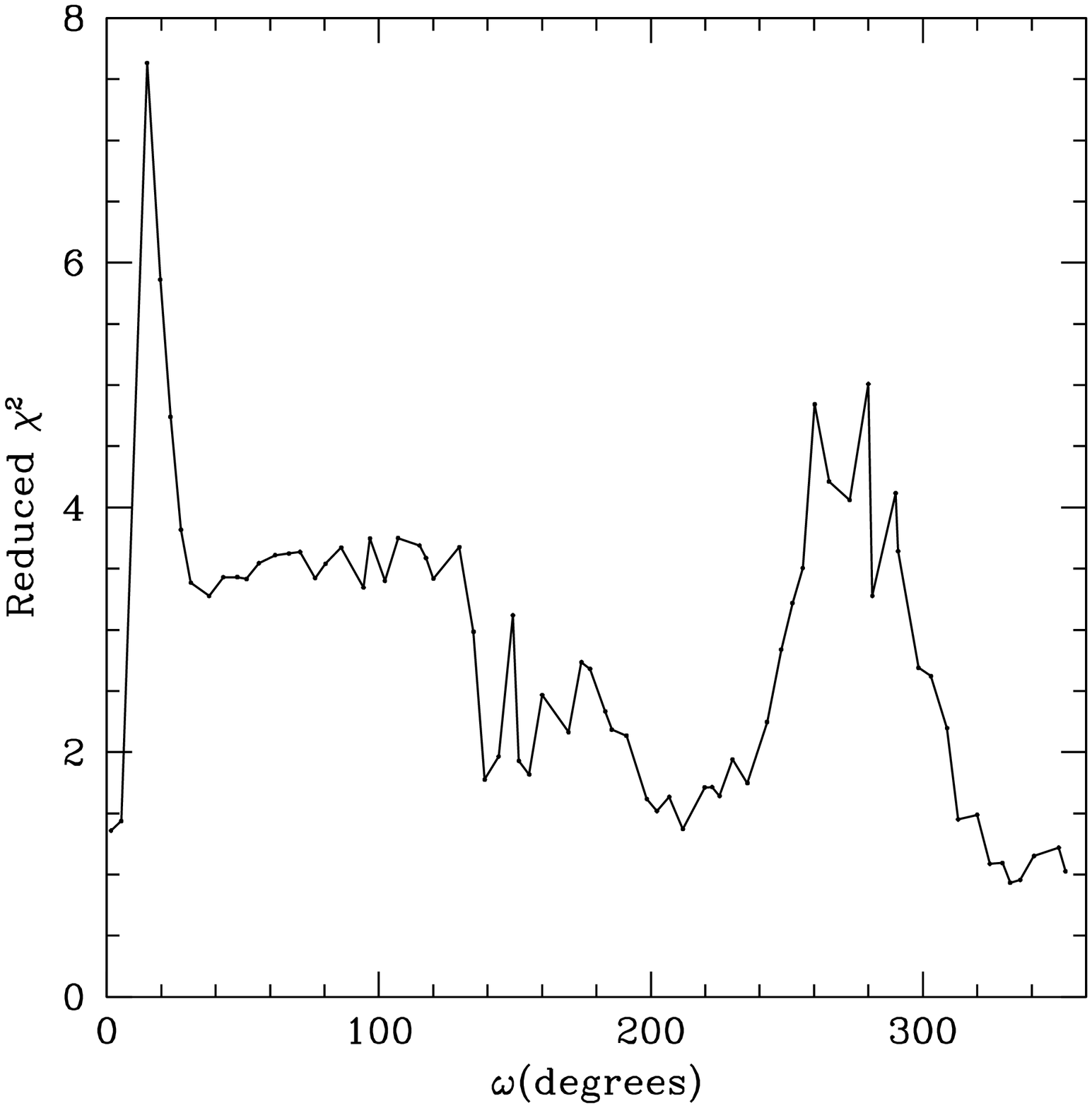}}
\caption{Results of genetic algorithm fitting for 2 planets (Model~B2). 
Left panel: Dependence of reduced $\chi^2$ on the outer planet's 
eccentricity.  Right panel: Same, but for the outer planet's argument of 
periastron ($\omega$).  This parameter displays two minima, and so is 
very poorly constrained by the available data. }
\label{chiB2}
\end{figure*}

Allowing the genetic algorithm to choose long periods for the second 
planet, we obtain a far better solution, with a reduced $\chi^2$ of 
0.80.  The parameters of this fit are given in Table~\ref{planetparams} 
as Model~B2; this fit and a periodogram of its residuals are shown in 
Figure~\ref{B2fits}.  Both Models A2 and B2 support a long period for 
the second planet, so the short period case discussed briefly above is 
rejected.  Again, the fitting process failed when $e$ and $\omega$ for 
the outer planet were free parameters, so we fixed their values at the 
best-fit from the genetic run (100,000 iterations).  Uncertainty 
estimates are obtained from the plots in Figure~\ref{chiB2}; for 
$\omega$, the $\chi^2$ surface has two minima, and so no formal 
uncertainty is quoted -- this parameter is very poorly constrained due 
to the long period of the outer planet.

In \citet{qian11} and \citet{horner11}, it was suggested that a third, 
distant outer planet may be present in the HU~Aqr system.  However, in 
light of the results of the 2-planet fits given in this section, which 
feature reduced $\chi^2$ less than 1.0, we see no need to invoke 
additional bodies to adequately fit the available data.

In summary, our analysis of two slightly different versions of the 
HU~Aquarii data yields evidence for two planets: a moderately well 
constrained one at $P=4647-4688$d with $e\sim$0.2, and a somewhat more 
poorly constrained one at $P=7215-8377$d with a poorly constrained but 
non-zero eccentricity.  While the outer planet's period varies by $\sim 
1200$d between these two solutions, we note that this represents just a 
2-$\sigma$ difference, given the period uncertainties.  Hence, there is 
a long period outer signal present, even if its period is not well 
determined.

\section{Dynamical Analysis}


Following the results detailed above, the observational data yield two 
distinct 2-planet solutions which are essentially identical in terms of 
their goodness-of-fit criteria. In both solutions, the best-fit system 
parameters are significantly different from those given by 
\citet{qian11}, whose dynamical stability was studied in some detail in 
\citet{horner11}.  In that work, the authors argued that the extreme 
levels of dynamical instability displayed by the planetary system 
provided firm evidence that, at the very least, the true parameters of 
the planets in the system were greatly different to those obtained 
through the analysis of \citet{qian11}.

How do the new solutions for the proposed HU~Aqr planets stand up to the 
same test?  In order to closely examine the dynamical stability of the 
newly proposed solutions, we followed \citet{horner11} and 
\citet{marshall10} and performed highly detailed dynamical simulations 
of the potential planetary systems.  Such simulations serve as a 
critically important additional test, since the algorithms used to this 
point include no physics; rather, the Keplerian fitting methods are 
simply seeking a lowest-$\chi^2$ solution regardless of the physicality 
of the resulting system parameters.

As in \citet{horner11}, we used the Hybrid integrator within the 
\textit{N}-body dynamics package Mercury \citep{chambers99} to perform 
our integrations.  For the two scenarios in question (Model~A2 and 
Model~B2), we created 50625 test planetary systems.  In each case, we 
followed our earlier work, and kept the initial orbit of the innermost 
planet fixed at its nominal best fit value.  The initial orbit of the 
outermost planet was then varied systematically in semi-major axis $a$, 
eccentricity $e$ and mean anomaly $M$, such that a total of 50625 unique 
solutions were tested.  For our tests of each of the two models, 45 
initial values of $a$ were tested, spread evenly across the full $\pm 
3\sigma$ error range in that parameter.  Similarly, 45 initial values of 
$e$ were tested in each case, with 25 different $M$ being considered for 
each $a-e$ pairing.  For Model~A2, in which the $e$ of the outermost 
planet was unconstrained, we tested eccentricities ranging between 0.005 
and 0.995, whilst for Model~B2, the tested $e$ values were spread across 
the full $\pm 3\sigma$ errors given in Table~\ref{planetparams}.

As in \citet{horner11}, we followed the dynamical evolution of each test 
system for a period of 100 million years, and recorded the times at 
which either of the planets was removed from the system.  Planets were 
removed if they collided with one another, hit the central body, or 
reached a barycentric distance of 100~AU.

\begin{figure*}
\subfigure[]{
 \includegraphics[scale=0.35,angle=90]{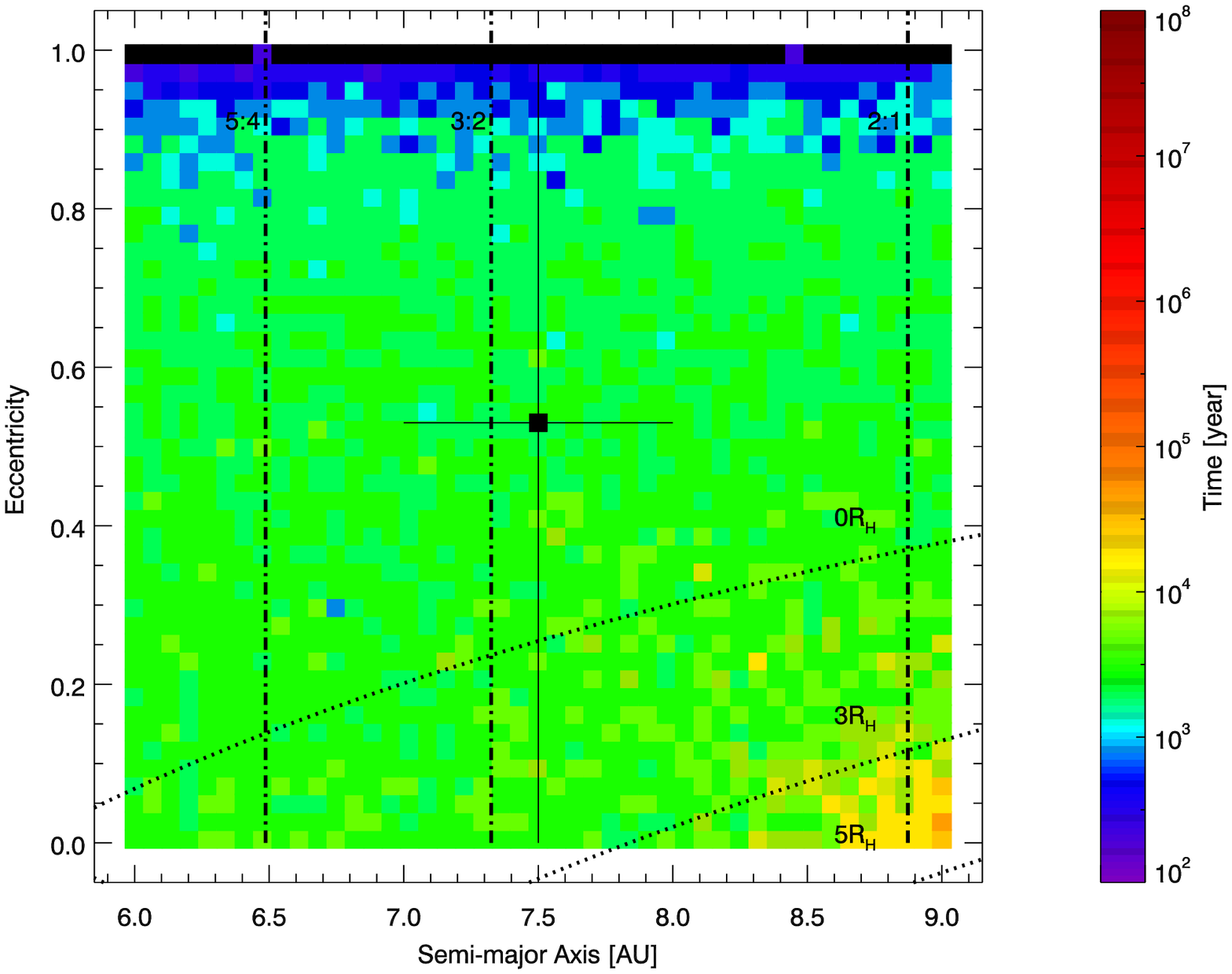}}
\subfigure[]{
 \includegraphics[scale=0.35,angle=90]{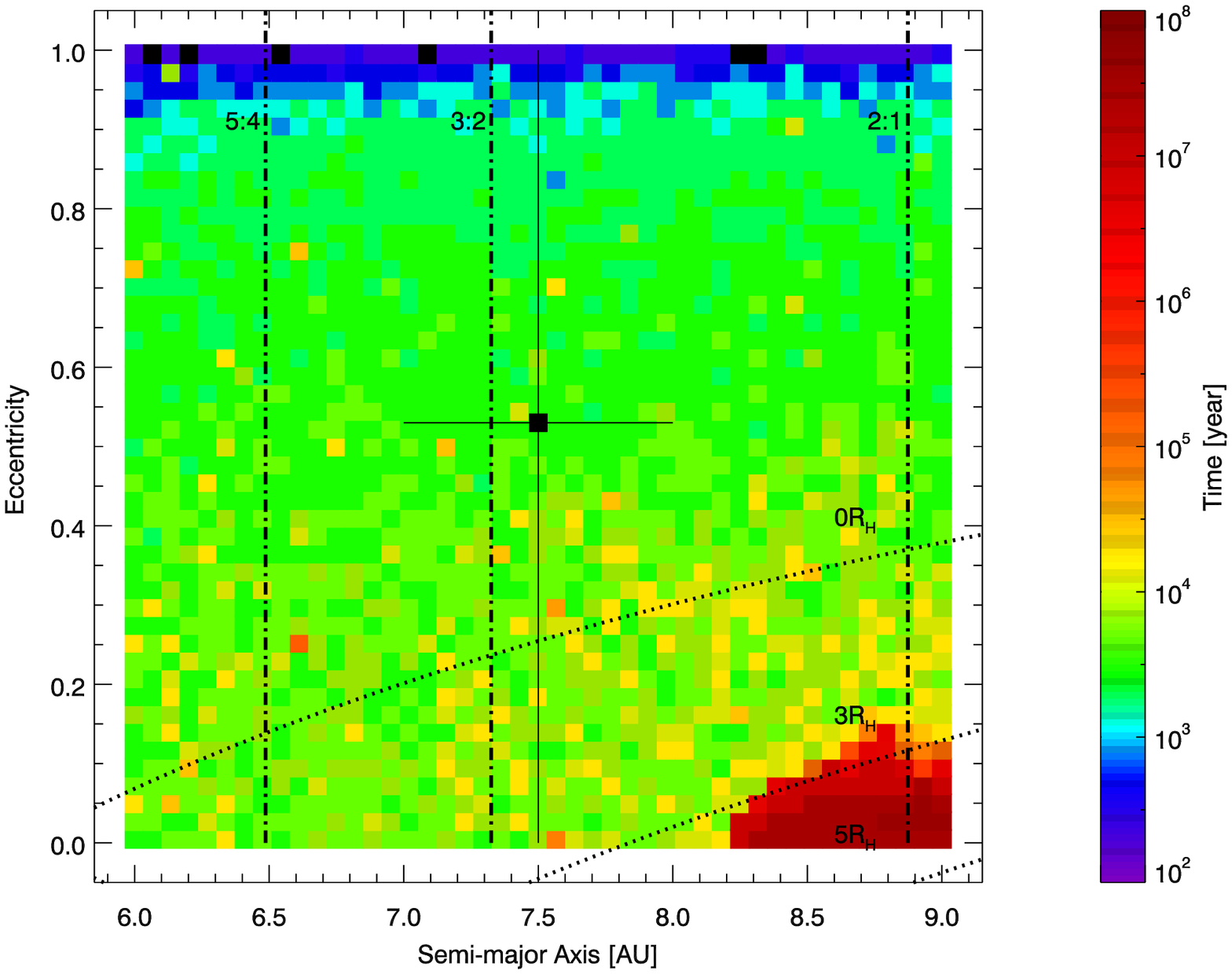}}
\caption{Left panel: Median lifetime of the outermost planet in the 
HU~Aqr system, based on the orbits fit by Model~A2, as a function of the 
planet's semi-major axis $a$, and eccentricity $e$.  The lifetime shown 
in each square is the median of the 25 distinct runs (each of which had 
the planet start at a different initial mean anomaly, $M$).  In both 
panels, the vertical dashed lines indicate relevant mean-motion 
resonances.  This panel reveals that the whole phase-space of potential 
orbits described in Model~A2 is highly dynamically unstable, with median 
lifetimes typically much less than $10^5$ years.  Right panel: Same as 
the left-hand panel, except that the mean, rather than the median, of 
the 25 initial mean anomaly values is calculated for each bin across the 
plot.  Whereas the left panel clearly showed the overall lack of 
stability of the plausible orbital solutions given by Model~A2, plotting 
the \textit{mean} lifetimes of each bin reveals that, for a small region 
of orbital space for which the outermost planet could approach no closer 
to the innermost planet than $3 R_{H}$, some fraction of the tested 
orbits were dynamically stable on much longer timescales.}
\label{A2dynamics}
\end{figure*}

\begin{figure*}
\subfigure[]{
 \includegraphics[scale=0.35,angle=90]{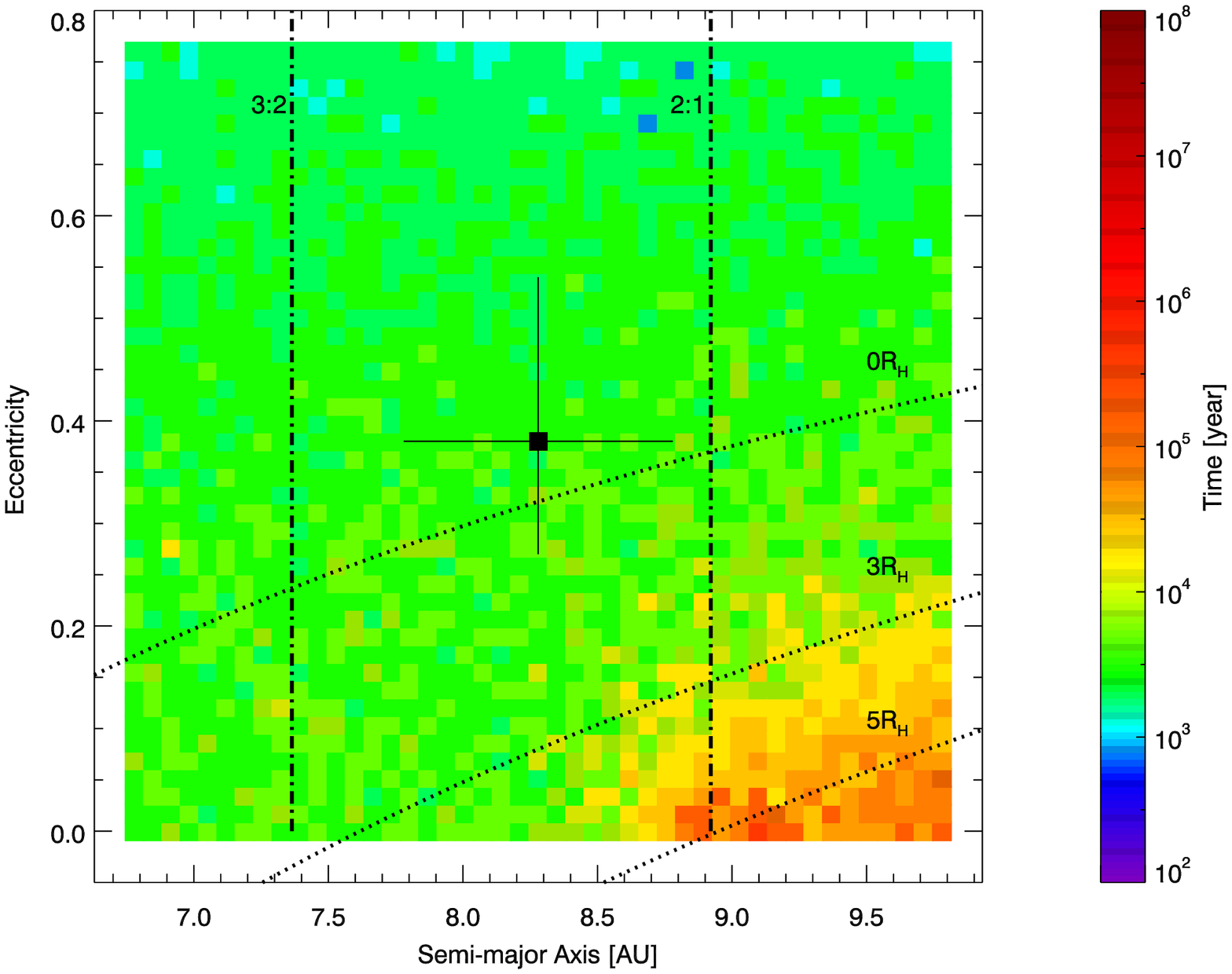}}
\subfigure[]{
 \includegraphics[scale=0.35,angle=90]{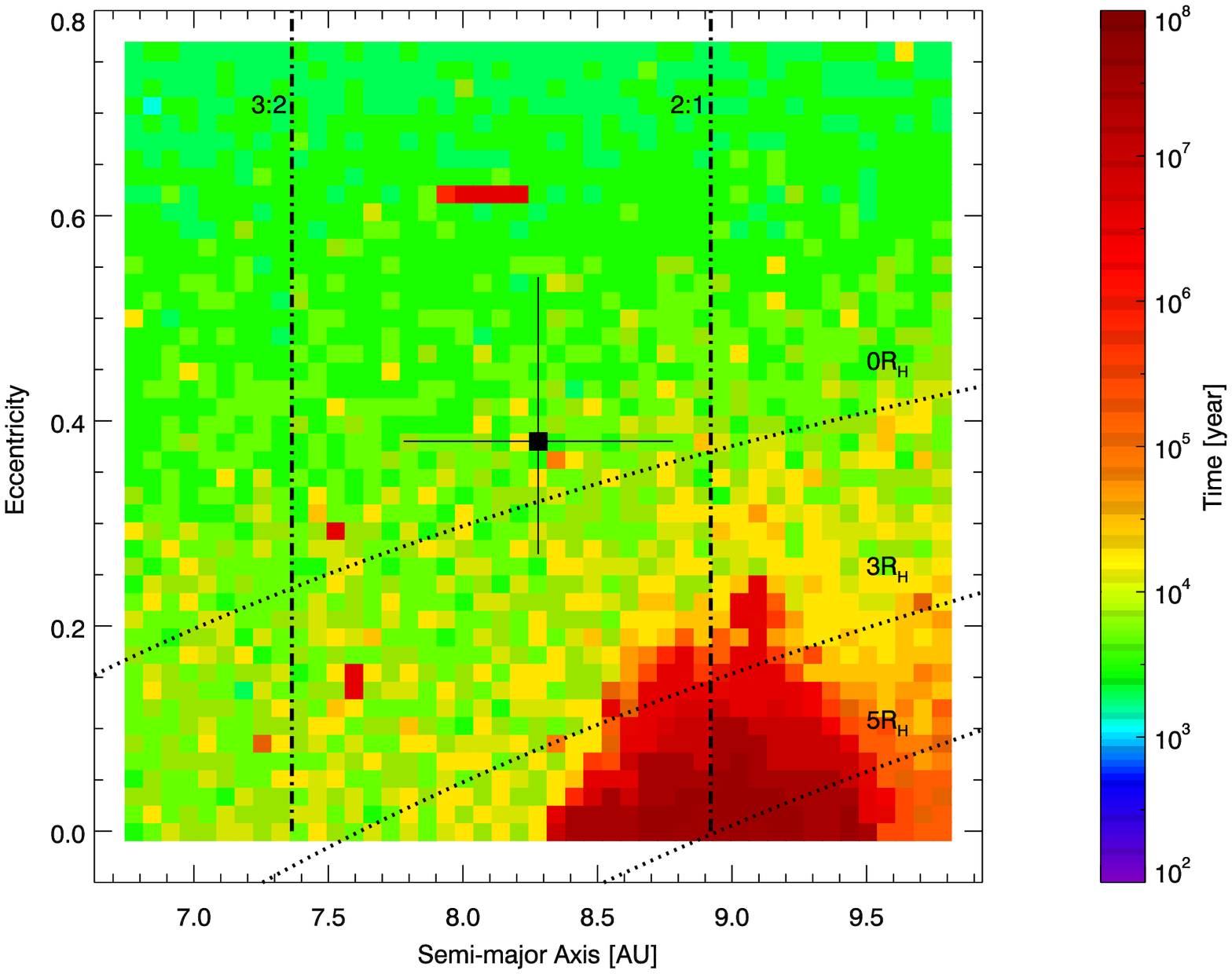}}
\caption{Left panel: Median lifetime of the outermost planet in the 
HU~Aqr system, based on the orbits fit by Model~B2, as a function of the 
planet's semi-major axis $a$, and eccentricity $e$, as described above.  
In both panels, the vertical dashed lines indicate relevant mean-motion 
resonances.  The great bulk of the tested $a-e$ phase space is highly 
dynamically unstable. Right panel: Same as the left-hand panel, except 
that the mean, rather than the median, of the 25 initial mean anomaly 
values is calculated for each bin across the plot.  As was the case for 
the results of Model~A2, this reveals that a small fraction of orbits 
tested can display long-term dynamical stability, though such orbits 
remain in the minority.}
\label{B2dynamics}
\end{figure*}

The results of our dynamical integrations can be seen in 
Figure~\ref{A2dynamics} and Figure~\ref{B2dynamics}.  The $3-\sigma$ 
region around the nominal orbits in $a-e$ space is clearly highly 
dynamically unstable.  As was seen in the dynamical study of the 
planetary system proposed by \citet{qian11}, low eccentricity orbital 
solutions which keep the planets separated at all times by at least 
three Hill radii offer some moderate increase in the potential stability 
of the system.  We note that these regions lie far from the central 
$1-\sigma$ of the error ellipse.  Furthermore, a fit with the outer 
planet's period and eccentricity fixed in this stable region has a 
reduced $\chi^2$ of 14.9, far worse than the best-fit results given in 
Table~\ref{planetparams}.

Much as was the case for the planetary system as proposed by 
\citet{qian11}, we therefore find that both of the best-fit models 
detailed above (Model~A2 and Model~B2) fail to stand up to dynamical 
scrutiny.  We find that the resulting dynamical instabilities make it 
exceedingly unlikely that the eclipse-timing variations observed for 
HU~Aqr are truly the result of perturbations from planetary mass objects 
in the system.  It seems reasonable, therefore, to examine more closely 
whether any other effects could cause timing variations of an 
appropriate periodicity and scale in such eclipsing polar systems.

\section{Alternatives to the Planet Hypothesis}

There are many causes of variability, spanning timescales from seconds 
to years, which are inherent to the nature of CVs. All of these 
mechanisms may have an observable impact on the variation of the $O-C$ 
curve, which may act to dilute, obscure, or mimic a signal that would 
otherwise be attributable to the presence of exoplanet(s).

At the time scale considered here (i.e.~thousands of days) the most 
likely source of the observed variation in the $O-C$ curve is the 
secondary star in the system, the M~dwarf.  With a rotation period on 
the order of a few hours (the result of it being tidally locked in its 
rotation about the primary, the white dwarf), the dynamo effect in the 
secondary could be large.  Assuming that the stars in such systems have 
a magnetic cycle similar to the Sun's double-peaked 22-year cycle, then 
their angular momentum distribution will change over time.  This will 
have the effect of altering the shape of the secondary, which in turn 
affects the gravitational attraction between the primary and secondary 
-- and hence the orbital period \citep{warner88, app92}.

This effect has already been observed in several CVs. Examples include 
U~Gem, which displays a $\sim$1 minute variation in orbital period over 
an eight year time scale \citep{eason83}, EX~Dra, whose period varies by 
1.2 minutes over approximately four years \citep{bap00}, and EX~Hya, 
whose period is modulated on a timescale of approximately 17.5 years 
\citep{hellier92}.

On the question of planetary survivability, \cite{qian11} mentioned that 
a gas giant planet could survive the planetary nebula stage, so long as 
it was located beyond orbital distances of about 3~AU \citep{vl07, 
kunitomo11}.  However, to make it to the planetary nebula phase, the 
planet must first survive the Asymptotic Giant Branch (AGB) phase.  A 
normal AGB star with a main sequence mass of 5~M$_{\odot}$ (extrapolated 
from the white dwarf mass of 0.88~M$_{\odot}$) is expected to reach a 
radius of 5.25~AU, enveloping both of the postulated planets from 
\cite{qian11}.  A planet entering the envelope of an AGB star would 
clearly have very little chance of surviving.  Planets orbiting 
post-main-sequence stars are thought to be either survivors of the 
planetary nebula or supernova phase \citep{colgate70, postnov92, 
veras11}, or formed from a second phase of planet formation, accreting 
from some of the material shed by the primary star \citep{tavani92, 
phinney93, hansen09}.  This ``second-generation'' planet formation 
scenario was explored after the discovery of the terrestrial-mass 
planets orbiting pulsars \citep{wf92}.  However, the mechanism by which 
Jupiter-mass planets could form in such an evolved system is not 
currently clear.

\section{Summary and Conclusions}

We have revisited the work of \citet{qian11}, who reported that timing 
variations in the mutual eclipses between the primary and secondary 
components of the HU~Aqr system were evidence that there were at least 
two Jupiter-mass planetary companions orbiting around the two central 
stars.

In this work, we applied the key tools used in the detection of planets 
around main sequence stars through the radial velocity technique to the 
data used by \citet{qian11}.  Our analysis resulted in two distinct 
two-planet fits which could, in theory, explain the observed timing 
variations.

Our derived orbital solutions differ significantly from those presented 
in that earlier work, but still fall prey to the same dynamical 
drawback. Simply put, the planets necessary in order to explain the 
eclipse-timing variations prove dynamically unstable on timescales so 
short as to seem unfeasible.

Our results therefore suggest that some other mechanism must instead be 
invoked in order to explain the observed variations.  The most likely 
candidate, based on earlier studies of eclipsing polar systems, is that 
the observed variations are the result of the interaction between the 
magnetic fields of the stars in the course of their stellar cycles.  
Such variations have been observed in a number of eclipsing polar 
systems in the past, and would be expected to yield variations on the 
scale observed, over similar timescales.

In light of these results, it would seem prudent in future to consider 
such behaviour as a potential source of signals that could mimic the 
presence of planets orbiting cataclysmic variable stars.  Rigorous 
dynamical testing of any planetary systems resulting from the analysis 
of transit timing data for such stars should become a key component of 
the analytical process.  These tests will prove critical in 
distinguishing between CVs which might plausibly host such interesting 
planetary systems, and those in which such planets are all but 
impossible.

\section*{Acknowledgments}

JH and CGT gratefully acknowledge the financial support of the 
Australian government through ARC Grant DP0774000.  RW is supported by a 
UNSW Vice-Chancellor's Fellowship.  JPM is partly supported by Spanish 
grant AYA 2008/01727, thanks Eva Villaver for constructive discussions 
of planet survivability, and gratefully acknowledges Maria Cunningham 
for funding his collaborative visit to UNSW.


\label{lastpage}



\begin{table}
  \centering
  \caption{Eclipse Timing Residuals for HU Aqr}
  \begin{tabular}{lll}
  \hline
JD-2400000 & O-C (s) & Uncertainty (s) \\
 \hline
49217.436369  &    -22.2  &    1.0  \\
49217.523190  &    -23.1  &    1.0  \\
49217.610010  &    -22.6  &    1.0  \\
49217.696830  &    -19.8  &    2.0  \\
49218.651855  &    -21.2  &    2.0  \\
49218.738675  &    -20.4  &    2.0  \\
49221.603749  &    -21.4  &    2.0  \\
49221.690569  &    -21.6  &    2.0  \\
49221.777389  &    -20.9  &    2.0  \\
50325.959283  &     -2.5  &    2.4  \\
50326.046103  &     -2.5  &    2.4  \\
50338.895523  &     -3.7  &    2.4  \\
50285.414154  &     -4.3  &   10.0  \\
50285.500975  &     -8.3  &   10.0  \\
50286.455999  &     -9.4  &   10.0  \\
50328.390254  &      6.0  &   10.0  \\
50328.477074  &      7.6  &   10.0  \\
50330.387123  &     -7.1  &   10.0  \\
50330.473944  &     -6.7  &   10.0  \\
51081.383615  &      8.3  &   10.0  \\
51481.278394  &     20.9  &    2.0  \\
51703.625447  &     22.3  &    7.0  \\
51704.580472  &     20.1  &    7.0  \\
52145.367661  &     22.8  &    4.0  \\
51703.625447  &     22.3  &    8.0  \\
51704.580472  &     20.1  &    8.0  \\
51821.440735  &     24.7  &    1.0  \\
51821.527555  &     24.7  &    1.0  \\
51350.874147  &     15.1  &    1.0  \\
51353.826041  &     15.4  &    1.0  \\
51354.867886  &     16.6  &    1.0  \\
51731.494797  &     23.5  &    1.0  \\
51731.581617  &     23.7  &    1.0  \\
52174.278855  &     20.9  &    1.0  \\
52174.365676  &     21.1  &    1.0  \\
52411.559018  &     14.9  &    1.0  \\
52552.381713  &     11.3  &    1.0  \\
52553.336737  &      8.9  &    1.0  \\
52787.665007  &      3.0  &    1.0  \\
52789.575056  &      1.7  &    1.0  \\
53205.444789  &     -6.8  &    1.5  \\
53205.531609  &     -7.1  &    1.5  \\
53209.525348  &     -6.0  &    1.5  \\
53212.564062  &     -5.7  &    1.5  \\
53293.307038  &     -7.6  &    1.5  \\
53295.303907  &     -9.2  &    1.5  \\
53296.258931  &     -7.5  &    1.5  \\
53299.297645  &     -7.6  &    1.5  \\
53533.539094  &     -6.8  &    1.5  \\
53918.500764  &     -4.8  &    1.5  \\
53925.446396  &     -4.9  &    1.5  \\
54320.479233  &      1.5  &    1.5  \\
54320.566053  &     -0.3  &    1.5  \\
52411.211737  &     11.6  &    5.0  \\
52779.937991  &      3.7  &    5.0  \\
52411.385378  &     16.8  &    2.0  \\
52411.385378  &     17.3  &    2.0  \\
52411.472198  &     16.8  &    2.0  \\
52413.642708  &     12.3  &    1.0  \\
52414.684553  &     16.2  &    1.0  \\
52783.671268  &      2.6  &    1.0  \\
53504.888361  &     -5.8  &    0.5  \\
53505.843386  &     -5.9  &    0.5  \\
53505.930206  &     -5.8  &    0.5  \\
53506.798410  &     -5.9  &    0.5  \\
53506.885230  &     -5.9  &    0.5  \\
53507.927075  &     -5.8  &    0.5  \\
54270.817962  &      2.4  &    1.4  \\
54270.904782  &      2.6  &    1.4  \\
54273.769856  &      2.5  &    1.4  \\
54273.856676  &      2.4  &    1.4  \\
54972.326823  &     -2.9  &    3.4  \\
55086.061552  &     -6.5  &    3.4  \\
55087.103397  &     -4.1  &    3.4  \\
55122.005199  &     -4.7  &    3.4  \\
55136.070104  &     -4.1  &    3.4  \\
55162.984429  &     -7.5  &    3.4  \\
55164.026274  &     -7.0  &    3.4  \\
55164.981299  &     -9.0  &    3.4  \\
55172.968776  &     -9.0  &    3.4  \\
55335.322931  &    -14.6  &    3.4  \\
 \hline
 \end{tabular}
\label{timings}
\end{table}


\begin{thebibliography}{99}

\bibitem[Applegate(1992)]{app92} Applegate, J.~H.\ 1992, ApJ, 385, 621

\bibitem[Baptista et al.(2000)]{bap00} Baptista, R., Catal{\'a}n, M.~S., 
\& Costa, L.\ 2000, MNRAS, 316, 529

\bibitem[Beuermann et al.(2010)]{b10} Beuermann, K., et 
al.\ 2010, A\&A, 521, L60

\bibitem[Chambers (1999)]{chambers99} Chambers, J. E.\ 1999, MNRAS, 304, 
793

\bibitem[Cochran et al.(2007)]{cochran07} Cochran, W.~D., Endl, 
M., Wittenmyer, R.~A., \& Bean, J.~L.\ 2007, ApJ, 665, 1407 

\bibitem[Colgate(1970)]{colgate70} Colgate, S.~A.\ 1970, Nature, 225, 
247

\bibitem[Eason et al.(1983)]{eason83} Eason, E.~L.~E., Africano, J.~L., 
Klimke, A., Worden, S.~P., Quigley, R.~J., \& Rogers, W.\ 1983, PASP, 
95, 58

\bibitem[Hansen et al.(2009)]{hansen09} Hansen, B.~M.~S., Shih, H.-Y., 
\& Currie, T.\ 2009, ApJ, 691, 382

\bibitem[Hellier(2001)]{hellier01} Hellier, C.\ 2001, Cataclysmic 
Variable Stars, Springer, 2001,

\bibitem[Hellier \& Sproats(1992)]{hellier92} Hellier, C., \& Sproats, 
L.~N.\ 1992, Information Bulletin on Variable Stars, 3724, 1

\bibitem[Horner et al.(2011)]{horner11} Horner, J., Marshall, J.~P., 
Wittenmyer, R.~A., \& Tinney, C.~G.\ 2011, MNRAS, L280

\bibitem[Jefferys et al.(1987)]{jeffreys87} Jefferys, W.~H.,
Fitzpatrick, M.~J., \& McArthur, B.~E.\ 1987, Celestial Mechanics, 41, 39

\bibitem[Kunitomo et al.(2011)]{kunitomo11} Kunitomo, M., Ikoma, M., 
Sato, B., Katsuta, Y., \& Ida, S.\ 2011, ApJ, 737, 66

\bibitem[K{\" u}rster et al.(1997)]{kurster97} K{\" u}rster, M., 
Schmitt, J.~H.~M.~M., Cutispoto, G., \& Dennerl, K.\ 1997, A\&A, 320, 
831

\bibitem[Lomb(1976)]{lomb76} Lomb, N.~R.\ 1976, Ap\&SS, 39, 447

\bibitem[Marshall et al.(2010)]{marshall10} Marshall, J., Horner, J., \& 
Carter, A.\ 2010, International Journal of Astrobiology, 9, 259

\bibitem[Phinney \& Hansen(1993)]{phinney93} Phinney, E.~S., 
\& Hansen, B.~M.~S.\ 1993, Planets Around Pulsars, 36, 371

\bibitem[Postnov \& Prokhorov(1992)]{postnov92} Postnov, 
K.~A., \& Prokhorov, M.~E.\ 1992, A\&A, 258, L17

\bibitem[Potter et al.(2011)]{potter11} Potter, S.~B., et al.\ 2011, 
arXiv:1106.1404

\bibitem[Qian et al.(2010)]{qian10} Qian, S.-B., Liao, W.-P., Zhu, 
L.-Y., \& Dai, Z.-B.\ 2010, ApJL, 708, L66

\bibitem[Qian et al.(2011)]{qian11} Qian, S.-B., et al.\ 2011, MNRAS, 
414, L16

\bibitem[Scargle(1982)]{scargle82} Scargle, J.~D.\ 1982, ApJ, 263, 835

\bibitem[Schwarz et al.(2009)]{schwarz09} Schwarz, R., 
Schwope, A.~D., Vogel, J., Dhillon, V.~S., Marsh, T.~R., Copperwheat, 
C., Littlefair, S.~P., \& Kanbach, G.\ 2009, A\&A, 496, 833

\bibitem[Tavani \& Brookshaw(1992)]{tavani92} Tavani, M., \& 
Brookshaw, L.\ 1992, Nature, 356, 320

\bibitem[Tinney et al.(2011)]{tinney11} Tinney, C.~G., Wittenmyer, 
R.~A., Butler, R.~P., Jones, H.~R.~A., O'Toole, S.~J., Bailey, J.~A., 
Carter, B.~D., \& Horner, J.\ 2011, ApJ, 732, 31

\bibitem[Veras et al.(2011)]{veras11} Veras, D., Wyatt, M.~C., Mustill, 
A.~J., Bonsor, A., \& Eldridge, J.~J.\ 2011, MNRAS, 1332

\bibitem[Villaver \& Livio(2007)]{vl07} Villaver, E., \& 
Livio, M.\ 2007, ApJ, 661, 1192

\bibitem[Warner(1988)]{warner88} Warner, B.\ 1988, Nature, 336, 129

\bibitem[Wolszczan \& Frail(1992)]{wf92} Wolszczan, A., \& Frail, D.~A.\ 
1992, Nature, 355, 145

\end{thebibliography}
\end{document}